\documentclass[aps, prd, superscriptaddress, nofootinbib, preprint]{revtex4-1}

\usepackage{bm,amssymb,slashed,graphicx,multirow,soul,mathtools,xspace,array}

\usepackage{hyperref}

\newcommand{\bbar}{\bar{b}}
\newcommand{\cbar}{\bar{c}}

\newcommand{\RL}{R(\Lambda_c)}
\newcommand{\Bbar}{\,\overline{\!B}}
\def\B0bar{\Bbar{}^0}

\def\LbarL{\bar\Lambda_\Lambda}
\def\epsc{\varepsilon_c}
\def\epsb{\varepsilon_b}

\newcommand{\mn}{{\mu\nu}}
\newcommand{\g}{\gamma}

\def\spnt{1}
\if\spnt1
	\newcommand{\up}{+}
	\newcommand{\dn}{-}
\else
	\newcommand{\up}{2}
	\newcommand{\dn}{1}
\fi	
\newcommand{\ov}{\overline}
\newcommand{\aS}{\alpha_s}
\newcommand{\haS}{{\hat{\alpha}_s}}
\newcommand{\nn}{\nonumber}
\newcommand{\GeV}{\text{GeV}}
\newcommand{\MeV}{\text{MeV}}
	
\def\lqcd{\Lambda_\text{QCD}}

\newcommand{\beq}{\begin{equation}}
\newcommand{\eeq}{\end{equation}}
\newcommand{\beqa}{\begin{eqnarray}}
\newcommand{\eeqa}{\end{eqnarray}}

\newcommand{\alSL}{\alpha_L^S}
\newcommand{\alSR}{\alpha_R^S}
\newcommand{\alVL}{\alpha_L^V}
\newcommand{\alVR}{\alpha_R^V}
\newcommand{\alTL}{\alpha_L^T}
\newcommand{\alTR}{\alpha_R^T}
\newcommand{\beSL}{\beta_L^S}
\newcommand{\beSR}{\beta_R^S}
\newcommand{\beVL}{\beta_L^V}
\newcommand{\beVR}{\beta_R^V}
\newcommand{\beTL}{\beta_L^T}
\newcommand{\beTR}{\beta_R^T}
\newcommand{\rC}{r_\Lambda}

\def\d{d}
\newcommand{\rl}{\rho_\ell}
\newcommand{\rt}{r_\ell}
\newcommand{\mSqq}{\hat q^2}
\newcommand{\WpmP}{\Sigma_+}
\newcommand{\WpmM}{\Sigma_-}
\newcommand{\thtau}{\theta_{\ell}}
\newcommand{\phtau}{\phi_{\ell}}

\newcommand{\phL}{\phi_{\Lambda}}

\newcommand{\Ot}{\Omega_{\times}}
\newcommand{\Op}{\Omega_{+}}
\newcommand{\RCpp}{R_{+ +}}
\newcommand{\RCpm}{R_{+ -}}
\newcommand{\RCmp}{R_{- +}}
\newcommand{\RCmm}{R_{- -}}

\makeatletter
\g@addto@macro\bfseries{\boldmath}
\makeatother

\allowdisplaybreaks
\let\origfootnote\footnote
\renewcommand{\footnote}[1]{%
   \begingroup
   \renewcommand{\baselinestretch}{1}%
   \origfootnote{#1}%
   \endgroup}
\begin{document}

\title{Precise predictions for $\Lambda_b\to\Lambda_c$ semileptonic decays}

\author{Florian U.\ Bernlochner}
\affiliation{Karlsruher Institute of Technology, 76131 Karlsruhe, Germany\\[4pt]}

\author{Zoltan Ligeti}
\affiliation{Ernest Orlando Lawrence Berkeley National Laboratory, 
University of California, Berkeley, CA 94720, USA\\[4pt]}

\author{Dean J.\ Robinson}
\affiliation{Ernest Orlando Lawrence Berkeley National Laboratory, 
University of California, Berkeley, CA 94720, USA\\[4pt]}
\affiliation{Santa Cruz Institute for Particle Physics and
Department of Physics, University of California Santa Cruz,
Santa Cruz, CA 95064, USA
\vspace*{.5cm}}

\author{William L.\ Sutcliffe} 
\affiliation{Karlsruher Institute of Technology, 76131 Karlsruhe, Germany\\[4pt]}

\begin{abstract}

We calculate the $\Lambda_b \to \Lambda_c \ell \nu$ form factors and decay
rates for all possible $b\to c
\ell\bar\nu$ four-Fermi interactions beyond the Standard Model, 
including nonzero charged lepton masses and terms up to order $\alpha_s\,
\Lambda_\text{QCD}/m_{c,b}$ and $\Lambda_\text{QCD}^2/m_c^2$ in the heavy quark effective theory. At this
order, we obtain model independent predictions for semileptonic $\Lambda_b \to
\Lambda_c$ decays in terms of only two unknown sub-subleading Isgur-Wise
functions, which can be determined from fitting LHCb and lattice QCD data.  
We thus obtain model independent results for $\Lambda_b\to
\Lambda_c\ell\bar\nu$ decays, including predictions for the ratio $R(\Lambda_c) = {\cal
B}(\Lambda_b\to \Lambda_c \tau\bar\nu) / {\cal B}(\Lambda_b\to \Lambda_c
\mu\bar\nu)$ in the presence of new physics, that are more
precise than prior results in the literature, and systematically
improvable with better data on the decays with $\mu$ (or $e$) in the final
state. We also explore tests of factorization in  $\Lambda_b \to \Lambda_c\pi$
decays, and emphasize the importance of measuring at LHCb the double
differential rate $d^2\Gamma(\Lambda_b\to\Lambda_c\ell\bar\nu) / (d q^2\,
d\cos\theta)$, in addition to the $q^2$ spectrum.

\end{abstract}

\maketitle

\tableofcontents
\clearpage

\section{Introduction}

In a recent paper~\cite{Bernlochner:2018kxh}, it was shown that LHCb data for
the semileptonic  $\Lambda_b \to \Lambda_c \mu \nu$ decays~\cite{Aaij:2017svr}
combined with lattice QCD calculations~\cite{Detmold:2015aaa}, provide
sensitivity for the first time to sub-subleading $\mathcal{O}(\lqcd^2/m_c^2)$
terms in the heavy quark effective theory (HQET) expansion~\cite{Georgi:1990um,
Eichten:1989zv} of the $\Lambda_b \to \Lambda_c$ semileptonic decay form
factors, independent of $|V_{cb}|$.  The $\mathcal{O}(\lqcd^2/m_c^2)$
corrections were found to have their expected characteristic size, suggesting
that the expansion in $\lqcd/m_c$ for baryon form factors is well-behaved up to
$\lqcd^2/m_c^2$ terms.  The same framework also resulted in a new standard model
(SM) prediction for the ratio
\begin{equation}
\label{RLdef}
\RL = \frac{\Gamma(\Lambda_b\to \Lambda_c\tau\bar\nu)}
  	{\Gamma(\Lambda_b\to \Lambda_c \mu\bar\nu)} = 0.324 \pm 0.004 \,,
\end{equation}
which is significantly more precise than prior results~\cite{Detmold:2015aaa,
Woloshyn:2014hka, Shivashankara:2015cta, Gutsche:2015mxa, Dutta:2015ueb, Azizi:2018axf,
DiSalvo:2018ngq}.

The ratio in Eq.~\eqref{RLdef} is of particular interest in light of the
persistent hints of deviations from the SM, in the ratios
\begin{equation}\label{RXdef}
R(D^{(*)}) = \frac{\Gamma(B\to D^{(*)}\tau\bar\nu)}
  {\Gamma(B\to D^{(*)} l\bar\nu)}\,, \qquad l = \mu, e\,,
\end{equation}
at approximately the $4\sigma$ level, once the measurements for the $D$ and
$D^*$ final states are combined~\cite{Amhis:2016xyh}.  The $\Lambda_b \to
\Lambda_c \mu \bar\nu$ decays involve the same underlying $b \to c \tau \nu$ new
physics (NP) operators as  $B\to D^{(*)}\tau\bar\nu$, but the HQET expansion for
the ground-state baryon form factors is simpler than for mesons.  The ``brown
muck" surrounding the heavy quark is in a spin and isospin zero ground state. A
consequence of this is a simpler expansion of the form factors, in which the
$\mathcal{O}(\lqcd/m_{c,b}\,,\, \alpha_s\, \lqcd/m_{c,b})$ subleading
contributions are determined by the leading order Isgur-Wise function, reducing
the number of free parameters in the form factor fits, and thereby providing
sensitivity to $\mathcal{O}(\lqcd^2/m_c^2)$ terms.

The spread in the uncertainties quoted for theoretical predictions for $R(D^*)$
in the SM are largely due to different estimates of $\mathcal{O}(\lqcd^2/m_c^2)$
effects~\cite{Bernlochner:2017jka, Bigi:2017jbd, Jaiswal:2017rve}. The very same
hadronic matrix elements are also crucial to resolve tensions between inclusive
and exclusive determinations of $|V_{cb}|$~\cite{Bernlochner:2017jka,
Bigi:2017njr, Grinstein:2017nlq, Bernlochner:2017xyx, Lattice:2015rga,
BDsLatticeAllw, Kaneko:2018mcr, Bigi:2017jbd, Jaiswal:2017rve}. The abundant
sample of $\Lambda_b$ baryons produced at the LHC may therefore provide a
complementary and theoretically cleaner laboratory to study the behavior of the
heavy quark expansion, identify possible NP effects, and extract $|V_{cb}|$.

In this paper, we expand and generalize the study of
Ref.~\cite{Bernlochner:2018kxh} beyond the SM, to include all $b \to c \tau
\bar\nu$ four-Fermi  operators, including those containing
right-handed (sterile) neutrinos.  
We compute the relevant form factors including
$\mathcal{O}(\lqcd^2/m_c^2)$ terms, and compare the fit results of
Ref.~\cite{Bernlochner:2018kxh} to the lattice QCD determinations of not only
the three vector and three axial vector SM form factors, but also the four NP
tensor current form factors.   We further emphasize the importance of measuring
at LHCb the double differential rate
$\d^2\Gamma(\Lambda_b\to\Lambda_c\ell\bar\nu) / (\d q^2\, \d\cos\theta)$ in
addition to the $q^2$ spectrum, and also explore tests of factorization in
$\Lambda_b \to \Lambda_c\pi$ decay.

\section{HQET expansion of the form factors}
\label{sec:hqet}

\subsection{Form factor definitions}

We are interested in the $\Lambda_b\to \Lambda_c$ matrix elements of operators
with all possible Dirac structures, for which we choose the basis
\begin{align}\label{eqn:Odef}
  O_V &= \bar c\,\g_\mu\, b\,, & O_A &= \bar c\, \g_\mu\g_5\, b\,, & \nn\\*	
  O_S &= \bar c\, b\,, &  O_P & = \bar c\, \g_5\, b\,, & O_T = \bar c\, \sigma_{\mu\nu}\, b\,,
\end{align}
with $\sigma_{\mu\nu} = (i/2)\, [\g_\mu,\g_\nu]$.   As done in
Refs.~\cite{Leibovich:1997tu, Leibovich:1997em,
Bernlochner:2016bci,Bernlochner:2017jxt} for excited charm mesons, we use the
conventions $\text{Tr}[\g^\mu\g^\nu\g^\sigma\g^\rho\g^5] = -4i
\epsilon^{\mu\nu\rho\sigma}$, so that $\sigma^{\mu\nu} \g^5 \equiv
+(i/2)\epsilon^{\mu \nu \rho \sigma} \sigma_{\rho \sigma}$.  (This is the
opposite of the common convention in the $\Bbar \to D^{(*)}\ell\bar\nu$
literature, which typically chooses $\text{Tr}[\g^\mu\g^\nu\g^\sigma\g^\rho\g^5]
= +4i \epsilon^{\mu\nu\rho\sigma}$, so that $\sigma^{\mu\nu} \g^5 \equiv
-(i/2)\epsilon^{\mu \nu \rho \sigma} \sigma_{\rho \sigma}$.)

The semileptonic $\Lambda_b\to \Lambda_c\ell\bar\nu$ form factors in HQET are
conventionally defined for the SM currents as~\cite{Isgur:1990pm, Falk:1992ws,
Manohar:2000dt}
\begin{align}
\label{HQETffdef}
\langle \Lambda_c(p',s')| \bar c\gamma_\nu b |\Lambda_b(p,s)\rangle
  &= \bar u(p',s') \big[ f_1 \gamma_\mu + f_2 v_\mu + f_3 v'_\mu \big]
  u(p,s)\,, \nn\\*
\langle \Lambda_c(p',s')| \bar c\gamma_\nu\gamma_5 b |\Lambda_b(p,s)\rangle
  &= \bar u(p',s') \big[ g_1 \gamma_\mu + g_2 v_\mu + g_3 v'_\mu \big] 
  \gamma_5\, u(p,s)\,,
\end{align}
where $p = m_{\Lambda_b}v$, $p' = m_{\Lambda_c}v'$, and the $f_i$ and $g_i$ are
functions of $w = v \cdot v' = (m_{\Lambda_b}^2 + m_{\Lambda_c}^2 - q^2)/(2
m_{\Lambda_b} m_{\Lambda_c})$.  The spinors are normalized to $\bar u(p,s)
u(p,s) = 2m$.  We further define the NP form factors,
\begin{align}
\label{HQETffdefnew}
\langle \Lambda_c(p',s')| \bar c\, b |\Lambda_b(p,s)\rangle
  &= h_S\, \bar u(p',s')\, u(p,s)\,, \nn\\*
\langle \Lambda_c(p',s')| \bar c \gamma_5 b |\Lambda_b(p,s)\rangle
  &= h_P\, \bar u(p',s')\, \gamma_5\, u(p,s)\,, \nn\\*
\langle \Lambda_c(p',s')| \bar c\, \sigma_{\mu\nu}\, b |\Lambda_b(p,s)\rangle
  &= \bar u(p',s') \big[ h_1\, \sigma_{\mu\nu}
  + i\, h_2 (v_\mu \gamma_\nu - v_\nu \gamma_\mu)
  + i\, h_3 (v'_\mu \gamma_\nu - v'_\nu \gamma_\mu) \nn\\*
  & \qquad\quad + i\, h_4 (v_\mu v'_\nu - v_\nu v'_\mu) \big] u(p,s)\,.
\end{align}
In the definition of the NP tensor current, the conventions are chosen to
simplify the $\alpha_s$ corrections when expressed in terms of the standard
coefficient functions. 

In full QCD, the form factors of the SM currents were instead traditionally
defined as~\cite{Falk:1992ws},
\begin{align}
\label{QCDffdef}
\langle \Lambda_c(p',s')| \bar c\gamma_\mu b |\Lambda_b(p,s)\rangle
  &= \bar u(p',s') \big[ F_1\, \gamma_\mu - i F_2\, \sigma_{\mu\nu}\, q^\nu
  + F_3\, q_\mu \big] u(p,s)\,, \nn\\
\langle \Lambda_c(p',s')| \bar c\gamma_\mu\gamma_5 b |\Lambda_b(p,s)\rangle 
  &= \bar u(p',s') \big[ G_1\, \gamma_\mu - i G_2\, \sigma_{\mu\nu}\, q^\nu
  + G_3\, q_\mu \big] \gamma_5\, u(p,s)\,.
\end{align}
Our notation for the form factors follows Ref.~\cite{Manohar:2000dt}; the
notation of Ref.~\cite{Falk:1992ws} corresponds to an exchange of upper and
lowercase symbols, $F_i\leftrightarrow f_i$ and $G_i\leftrightarrow g_i$,
in Eqs.~\eqref{HQETffdef}  and~\eqref{QCDffdef}. The relations between the
form factors in Eqs.~\eqref{HQETffdef}  and~\eqref{QCDffdef} are given in the
Appendix~\ref{sec:FFapp}.

\subsection{Form factors in HQET}

The ground state baryons are singlets of heavy quark spin symmetry, because the
light degrees of freedom, the ``brown muck", are in the spin-0 state.  Hence,
the baryon masses can be written as
\begin{equation}\label{mass}
m_{\Lambda_Q} = m_Q + \LbarL - {\lambda_1^\Lambda \over 2 m_Q} 
  + \ldots \,, \qquad Q=b,c\,,
\end{equation}
where the ellipsis denote terms suppressed by more powers of $\Lambda_{\rm
QCD}/m_Q$.  The parameter $\LbarL$ is the energy of the light degrees of freedom
in the $m_Q\to\infty$ limit.  The $\lambda_1^\Lambda$ parameter is related to
the heavy quark kinetic energy in the $\Lambda$ baryon.  We use $m_{\Lambda_b} =
5.620\,$GeV, $m_{\Lambda_c} = 2.286$\,GeV~\cite{PDG}, and employ the $1S$ short
distance mass scheme~\cite{Hoang:1998ng, Hoang:1998hm, Hoang:1999ye} to
eliminate the leading renormalon ambiguities in the definition of the quark
masses and $\LbarL$.  Details of the $1S$ scheme treatment can be found in
Ref.~\cite{Bernlochner:2017jka}.  In particular, we treat $m_b^{1S} =
(4.71\pm0.05)\,\GeV$ and $\delta m_{bc} = m_b-m_c = (3.40\pm0.02)\,\GeV$ as
independent parameters~\cite{Ligeti:2014kia}.  (The latter is well constrained
by $B\to X_c\ell\bar\nu$ spectra~\cite{Bauer:2004ve, Bauer:2002sh}.)  We match
HQET onto QCD at scale $\mu = \sqrt{m_b m_c}$, so that $\aS \simeq 0.26$.  For
example, using Eq.~(\ref{mass}) for both $\Lambda_b$ and $\Lambda_c$ to
eliminate $\lambda_1^\Lambda$, at $\mathcal{O}(\aS)$ we obtain $\LbarL = (0.81
\pm 0.05)\,\GeV$ and $\lambda_1^\Lambda = - (0.24\pm0.08)\,\GeV^2$. (Similar
HQET-based discussions can be found for other decay modes, $B\to
D^{(*)}\ell\bar\nu$~\cite{Bernlochner:2017jka}, $B\to
D^{**}\ell\bar\nu$~\cite{Leibovich:1997tu, Leibovich:1997em,
Bernlochner:2016bci, Bernlochner:2017jxt}, and $\Lambda_b \to
\Lambda_c^*\ell\bar\nu$~\cite{Leibovich:1997az, Boer:2018vpx}.)

Making the transition to HQET~\cite{Georgi:1990um, Eichten:1989zv}, at leading
order in $\lqcd/m_{c,b}$,
\begin{equation}\label{leading}
\langle \Lambda_c(v',s')| \bar c\, \Gamma b\, |\Lambda_b(v,s)\rangle 
  = \zeta(w)\, \bar u(v',s')\, \Gamma\, u(v,s) \,,
\end{equation}
where $u(v,s)$ satisfies $\slashed{v}\, u(v,s) = u(v,s)$ and $\zeta(w)$ is the
Isgur-Wise function for ground state baryons~\cite{Isgur:1990pm}, satisfying
$\zeta(1)=1$. At leading order, one finds
\begin{align}
& f_1(w) = g_1(w) = h_S(w) =  h_P(w) = h_1(w) = \zeta(w)\,, \nn\\*
& f_2(w) = f_3(w) = g_2(w) = g_3(w) = h_2(w) = h_3(w) = h_4(w) = 0\,.
\end{align}

At order $\lqcd/m_{c,b}$ a remarkable simplification occurs compared to meson
decays. The $\mathcal{O}(\lqcd/m_{c,b})$ corrections from the matching of the
$\cbar\, \Gamma b$ heavy quark current onto HQET~\cite{Falk:1990yz, Falk:1990cz,
Neubert:1992qq} can be expressed in terms of $\LbarL$ and the leading order
Isgur-Wise function $\zeta(w)$~\cite{Georgi:1990ei}.  In addition, for
$\Lambda_b \to \Lambda_c$ transitions, i.e., between the ground state baryons,
there are no ${\cal O}(\lqcd/m_{c,b})$ contributions from the chromomagnetic
operator.  The kinetic energy operator in the ${\cal O}(\lqcd/m_{c,b})$ HQET
Lagrangian gives rise to  a heavy quark spin symmetry conserving subleading
term, parametrized by $\zeta_{\rm ke}(w)$, which can be absorbed into the
leading order Isgur-Wise function by redefining $\zeta$ via
\begin{equation}
	\label{eqn:KEabs}
	\zeta(w) +  (\epsc + \epsb)\, \zeta_{\rm ke}(w) \to \zeta(w)\,,
\end{equation}
where $\varepsilon_{c,b} = \LbarL/(2\, m_{c,b})$.  Luke's
theorem~\cite{Luke:1990eg} implies $\zeta_{\rm ke}(1) = 0$, so the normalization
$\zeta(1) = 1$ is preserved.   Thus, no additional unknown functions beyond
$\zeta(w)$ are needed to parametrize the ${\cal O}(\lqcd/m_{c,b})$ corrections.
Perturbative corrections to the heavy quark currents can be computed by matching
QCD onto HQET~\cite{Falk:1990yz, Falk:1990cz, Neubert:1992qq}, and introduce no
new hadronic parameters.  The same also holds for the order $\alpha_s\,
\lqcd/m_{c,b}$ corrections~\cite{Neubert:1993iv, Neubert:1993mb}. 

The $\mathcal{O}(\lqcd^2/m_{c,b}^2)$ corrections are parametrized by six linear
combinations of sub-subleading Isgur-Wise functions, $b_{1,\,
\ldots\,,\,6}$~\cite{Falk:1992ws}, which are functions of $w$. Only two
of these, $b_{1,2}(w)$, occur at $\mathcal{O}(\lqcd^2/m_c^2)$. The redefinition
in Eq.~\eqref{eqn:KEabs} introduces additional $\epsilon_c^2\,
\zeta_{\text{ke}}(w)$ terms, which can be reabsorbed into $b_{1,2}(w)$. We may
then define
\begin{equation}\label{eqn:hatHdef}
\big\{ \hat f_i(w)\,,\ \hat g_i(w)\,,\ \hat h_i(w),\ \hat b_i(w) \big\}
  = \big\{ f_i(w)\,,\ g_i(w)\,,\ h_i(w)\,,\ b_i(w) \big\} \big/ \zeta(w) \,.
\end{equation}
Thus, including $\alpha_s$, $\lqcd/m_{c,b}$, $\alpha_s\, \lqcd/m_{c,b}$, and
$\lqcd^2/m_c^2$ corrections, the SM form factors are~\cite{Bernlochner:2018kxh}
\begin{align}
\label{ffexpsm}
\hat f_1 &= 1 + \haS C_{V_1} + \epsc + \epsb + \haS \Big[ C_{V_1}
   + 2(w-1)C'_{V_1}\Big] (\epsc + \epsb)
  + \frac{\hat b_1-\hat b_2}{4m_c^2} + \ldots \,, \nn\\*
\hat f_2 & = \haS C_{V_2} - \frac{2\, \epsc}{w+1}
  + \haS\bigg[ C_{V_2} \frac{3w-1}{w+1} \epsb - 
  \big[2C_{V_1} - (w-1) C_{V_2} + 2C_{V_3}\big]  \frac{\epsc}{w+1} \nn\\*
   & \qquad + 2(w-1)C'_{V_2} (\epsc + \epsb)\bigg]
  + \frac{\hat b_2}{4m_c^2} + \ldots, \nn\\
\hat f_3 &= \haS C_{V_3} - \frac{2\, \epsb}{w+1} 
  + \haS\bigg[  C_{V_3} \frac{3w-1}{w+1} \epsc
  - \big[2C_{V_1} + 2C_{V_2} - (w-1) C_{V_3}\big] \frac{\epsb}{w+1} \nn\\
  & \qquad + 2(w-1) C'_{V_3} (\epsc + \epsb)\bigg] + \ldots \,, \nn\\
\hat g_1 &= 1 + \haS C_{A_1} + (\epsc + \epsb)\, \frac{w-1}{w+1} 
  + \haS\bigg[ C_{A_1}\, \frac{w-1}{w+1}
  + 2(w-1)C'_{A_1}\bigg] ( \epsc + \epsb )
  + \frac{\hat b_1}{4m_c^2} + \ldots \,, \nn\\
\hat g_2 &= \haS C_{A_2} - \frac{2\, \epsc}{w+1}
  + \haS \bigg[C_{A_2} \frac{3w+1}{w+1} \epsb - \big[2C_{A_1} - (w+1) C_{A_2} + 2C_{A_3}\big]
  \frac{\epsc}{w+1} \nn\\*
  & \qquad + 2(w-1)C'_{A_2} (\epsc + \epsb)\bigg]
  + \frac{\hat b_2}{4m_c^2} + \ldots , \nn\\
\hat g_3 &= \haS C_{A_3} + \frac{2\, \epsb}{w+1}
  + \haS\bigg[C_{A_3} \frac{3w+1}{w+1} \epsc
  +  \big[2C_{A_1} - 2C_{A_2} + (w+1) C_{A_3}\big] \frac{\epsb}{w+1} \nn\\*
  & \qquad + 2(w-1)C'_{A_3} (\epsc + \epsb)\bigg] + \ldots \,,
\end{align}
where the $C_{\Gamma_i}$ are functions of $w$, and $\haS = \aS/\pi$.  (We use
the notation of Ref.~\cite{Manohar:2000dt}; explicit expressions for
$C_{\Gamma_i}$ are in Ref.~\cite{Bernlochner:2017jka}.)  In Eq.~(\ref{ffexpsm}),
primes denote $\partial/\partial w$ and the ellipses denote ${\cal
O}(\epsc\epsb,\, \epsb^2,\, \epsc^3)$ and higher order terms in $\lqcd/m_Q$
and/or $\alpha_s$.  Equation~(\ref{ffexpsm}) agrees with Eq.~(4.75) in
Ref.~\cite{Neubert:1993mb} (where a redefinition different from
Eq.~\eqref{eqn:KEabs} was used).

For the expansions of the form factors parametrizing the BSM currents, we
obtain,
\begin{align}
\label{ffexpbsm}
\hat h_S &= 1 + \haS\, C_S + (\epsc + \epsb)\, \frac{w-1}{w+1}
  + \haS \bigg[ C_S \, \frac{w-1}{w+1} + 2(w-1)C'_S\bigg] (\epsc + \epsb)
  + \frac{\hat b_1}{4m_c^2} + \ldots \,,\nn\\
\hat h_P &= 1 + \haS\, C_P + \epsc + \epsb
  + \haS \bigg[ C_P + 2(w-1)C'_P \bigg](\epsc + \epsb)
  + \frac{\hat b_1-\hat b_2}{4m_c^2} + \ldots  \,,\nn\\
\hat h_1 & = 1 + \haS\, C_{T_1} + (\epsc + \epsb)\, \frac{w-1}{w+1}
  + \haS \bigg[ C_{T_1} \frac{w-1}{w+1}
  + 2(w-1)C'_{T_1} \bigg](\epsc + \epsb)
  + \frac{\hat b_1}{4m_c^2} + \ldots \,,\nn\\
\hat h_2 &= \haS\, C_{T_2} - \frac{2\, \epsc}{w+1}
  + \haS \bigg[ C_{T_2} \frac{3w+1}{w+1} \epsb - \big[2C_{T_1} - (w+1) C_{T_2} + 2C_{T_3}\big]
  \frac{\epsc}{w+1} \nn\\
  & \qquad + 2(w-1)C'_{T_2} (\epsc + \epsb)\bigg]
  + \frac{\hat b_2}{4m_c^2} + \ldots \,,\nn\\
\hat h_3 &= \haS\, C_{T_3} + \frac{2\, \epsb}{w+1}
  + \haS \bigg[ C_{T_3} \frac{3w+1}{w+1} \epsc
  +  \big[2C_{T_1} - 2C_{T_2} + (w+1) C_{T_3}\big] \frac{\epsb}{w+1} \nn\\
  & \qquad + 2(w-1)C'_{T_3} (\epsc + \epsb)\bigg]
  + \ldots \,, \nn\\
\hat h_4 &= \haS\, \frac2{w+1}\, \big( C_{T_3} \epsc -C_{T_2} \epsb \big)
  + \ldots \,.
\end{align}
Similar to $f_3$ and $g_3$, neither of the $h_3$ and $h_4$ form factors
receive $\lqcd^2/m_c^2$ corrections.  The structure of $h_{1,2,3}$ is
similar to $g_{1,2,3}$, while $h_4$ is non-zero only at
$\mathcal{O}(\aS\, \lqcd/m_{c,b})$.

\subsection{Differential decay rates and forward-backward asymmetry}

In Appendix~\ref{app:NPampl}, we collect explicit expressions for the $\Lambda_b
\to \Lambda_c \ell \nu$ amplitudes for all NP operators, including contributions
from massless right-handed sterile neutrinos~\cite{Asadi:2018wea, Greljo:2018ogz}.
Including the charged lepton mass dependence, and defining $\theta$ as the angle
between the lepton and the $\Lambda_c$ momentum in the dilepton rest
frame,\footnote{This angle is not measurable in the $\tau$ channel by present
experiments, because neither the $\Lambda_b$ nor $\tau$ momentum can be
precisely reconstructed.  In principle, if the $\Lambda_b$ momentum was known
and the $\tau \to 3\pi\nu$ decay mode was used to reconstruct the $\tau$ vertex,
then $\theta$ could be reconstructed.} the SM double differential decay rate is
\begin{align}
\label{eqn:dGdwdcosth}
\frac{\d^2\Gamma}{\d w\, \d\cos\theta} 
	& = \frac{G_F^2\, m_{\Lambda_b}^5 |V_{cb}|^2}{48\,\pi^3}\, 
	\frac{(\mSqq - \rl)^2}{\hat q^4}\, \rC^3\, \sqrt{w^2-1}\, 
	\bigg[ \bigg(1 + \frac{\rl}{2\mSqq}\bigg) \Big( {\cal H}_+ + 2\mSqq
	{\cal H}_1 \Big) + \frac{3\rl}{2\mSqq}\,{\cal H}_0 \nn\\
	& - 3\sqrt{w^2-1}\, \bigg(\! 2f_1\,g_1\, \mSqq 
	- \frac{\rho_\ell}{\mSqq}\,{\cal H}_{+0} \bigg) \cos\theta 
	+ \bigg(1 - \frac{\rl}{\mSqq}\bigg) \Big(\mSqq\, {\cal H}_1 - {\cal H}_+
	\Big) \frac{3\cos^2\theta - 1}2\, \bigg] ,
\end{align}
where $\rl = m_\ell^2/m_{\Lambda_b}^2$, $\rC = m_{\Lambda_c} /
m_{\Lambda_b}$, $\mSqq \equiv q^2/m_{\Lambda_b}^2 = 1 - 2\rC w + \rC^2$,
\begin{align}
	{\cal H}_1 & = (w-1) f_1^2 + (w+1)g_1^2\,, &  {\cal H}_+  & = (w-1){\cal F}_+^2 + (w+1){\cal G}_+^2\,,\nn\\
	{\cal H}_0 & =  (w+1) {\cal F}_0^2+ (w-1){\cal G}_0^2\,,  & {\cal H}_{+0}  & = {\cal F}_+{\cal F}_0 + {\cal G}_+{\cal G}_0\,,
\end{align}
and
\begin{align}\label{F0G0}
\mathcal{F}_+ & = (1+\rC)f_1 + (w+1)(\rC\, f_2+f_3)\,, \nn\\*
\mathcal{G}_+ & = (1-\rC)g_1 - (w-1)(\rC\, g_2+g_3)\,, \nn\\*
\mathcal{F}_0 & = (1-\rC)f_1 - (\rC w - 1) f_2 + (w-\rC)f_3\,, \nn\\* 
\mathcal{G}_0 & = (1 + \rC)g_1 + (\rC w - 1)g_2 - (w-\rC)g_3\,. 
\end{align}
The double differential rate in Eq.~\eqref{eqn:dGdwdcosth} can be at most a
degree-two polynomial in $\cos\theta$, and it was written in
Eq.~(\ref{eqn:dGdwdcosth}) in the Legendre polynomial basis, so that only the
zeroth order term in the first line contributes to the $d\Gamma/d q^2$, after
integration over $d\cos\theta$.

The single differential rate in the SM is correspondingly
\begin{equation}\label{dGdwSM}
\frac{\d\Gamma}{\d w} = 
  \frac{G_F^2\, m_{\Lambda_b}^5 |V_{cb}|^2}{24\,\pi^3}\,
  \frac{(\mSqq - \rl)^2}{\hat q^4}\, \rC^3\, \sqrt{w^2-1}\, \bigg[
  \bigg(1 + \frac{\rl}{2\mSqq}\bigg) \Big( {\cal H}_+ + 2\mSqq {\cal H}_1 \Big) 
  + \frac{3\rl}{2\mSqq}\,{\cal H}_0 \bigg] ,
\end{equation}
and the forward-backward asymmetry is given by
\begin{align}
	\label{eqn:AFB}
	\frac{\d A_{\rm FB}}{\d w} &= \bigg[\int_0^1 - \int_{-1}^0\bigg]\,
	\frac{\d^2\Gamma}{\d w\, \d\cos\theta}\, \d \cos \theta \nn\\*
	&= -\frac{G_F^2\, m_{\Lambda_b}^5 |V_{cb}|^2}{16\,\pi^3}\,
	\frac{(\mSqq - \rl)^2}{\hat q^4}\, \rC^3\ (w^2-1)\,
	\bigg(2f_1\, g_1\, \mSqq - \frac{\rho_\ell}{\mSqq}\, {\cal H}_{+0} \bigg)\,.
\end{align}
Our result in Eq.~\eqref{dGdwSM} agrees with those in Refs.~\cite{Boyd:1997kz,
Detmold:2015aaa}.  Including all possible NP current operators and a nonzero
charged lepton mass, our result for $\d\Gamma/\d w$ as derived from
Appendix~\ref{app:NPampl} agrees with the result for SM neutrinos in Eq.~(2.51)
of Ref.~\cite{Datta:2017aue}. We see from Eqs.~\eqref{eqn:dGdwdcosth} or
\eqref{eqn:AFB} that the $\theta$ distribution in the light lepton modes gives
sensitivity to the product $f_1\, g_1$, which is not present in $d\Gamma/dw$. 
The quadratic term in $\cos\theta$ in the angular distribution provides
sensitivity to the combination $\mSqq {\cal H}_1 - {\cal H}_+$. Thus, just like
in the case of $b\to s\ell^+\ell^-$~\cite{Lee:2006gs}, measuring the
dependencies on all three polynomials of $\cos\theta$,  gives information on
the form factors beyond measuring only $d\Gamma/d q^2$ and $dA_{\rm FB}/d q^2$.

To gain more information than obtainable from Eq.~\eqref{eqn:dGdwdcosth}, the
distribution of the $\Lambda_c$ decay products would have to be studied.  Such
an analysis would be simplest for two-body decays, such as $\Lambda_c \to \Lambda(p
\pi^-) \pi^+$~\cite{Shivashankara:2015cta}.  This channel loses an order of
magnitude in statistics compared to the commonly used $\Lambda_c \to p K \pi$
reconstruction, however, a model independent description of this three-body
decay amplitude is not currently available.  
With much higher statistics and using
$\Lambda_c \to \Lambda\pi^+$, the measurement of all $\Lambda_b\to \Lambda_c$
form factors would be similar to that for $\Lambda_c\to \Lambda
e\nu$~\cite{Korner:1991ph, Crawford:1995wz, Hinson:2004pj}, requiring measuring
distributions in three angles (as for $B\to (D^* \to D \pi) l \bar\nu$).

If NP only modifies the (axial)vector interactions (see e.g. Refs.~\cite{Shivashankara:2015cta,Dutta:2015ueb,Li:2016pdv} for other cases), which may be the most
plausible scenario, then Eqs.~\eqref{eqn:dGdwdcosth} -- \eqref{eqn:AFB} are
simply modified via the replacements
\begin{equation}
	f_i \to f_i(1 + g_L + g_R)\,, \qquad g_i \to g_i(1 + g_L - g_R)\,,
\end{equation}
and, in particular,
\begin{equation}
	\label{eqn:AFBmod}
	\frac{\d A_{\rm FB}}{\d w} \to \frac{\d A_{\rm FB}}{\d w}\, \big[(1+g_L)^2 -g_R)^2\big]\,.
\end{equation}
In the $m_l=0$ limit, i.e., in the $\Lambda_c\mu\nu$ and $\Lambda_c e\nu$
modes, the forward-backward asymmetry only receives further contributions from
tensor--(pseudo)scalar interference, even in the presence of arbitrary NP.  The
relation in Eq.~\eqref{eqn:AFBmod} is then valid in the light lepton modes, as
long as NP does not simultaneously generate (pseudo)scalar and tensor operators.

\section{Fits to LHC\lowercase{b} and lattice QCD data}

\subsection{SM form factor fits}
\label{sec:fitLHCbVA}

\begin{figure}[t]
\includegraphics[width=0.45\textwidth, clip, bb=10 0 420 310]{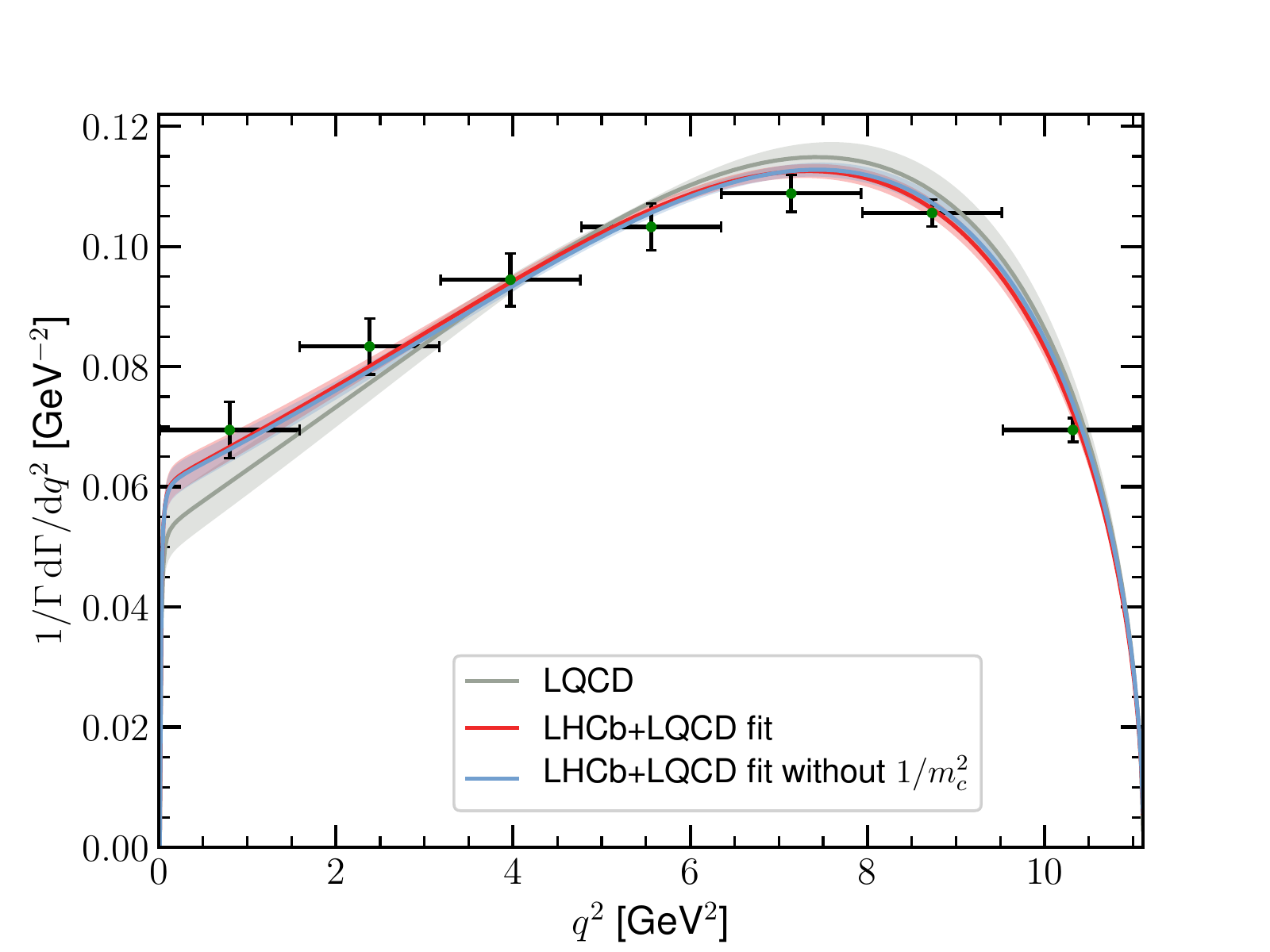}
\hfil
\includegraphics[width=0.45\textwidth, clip, bb=10 0 420 310]{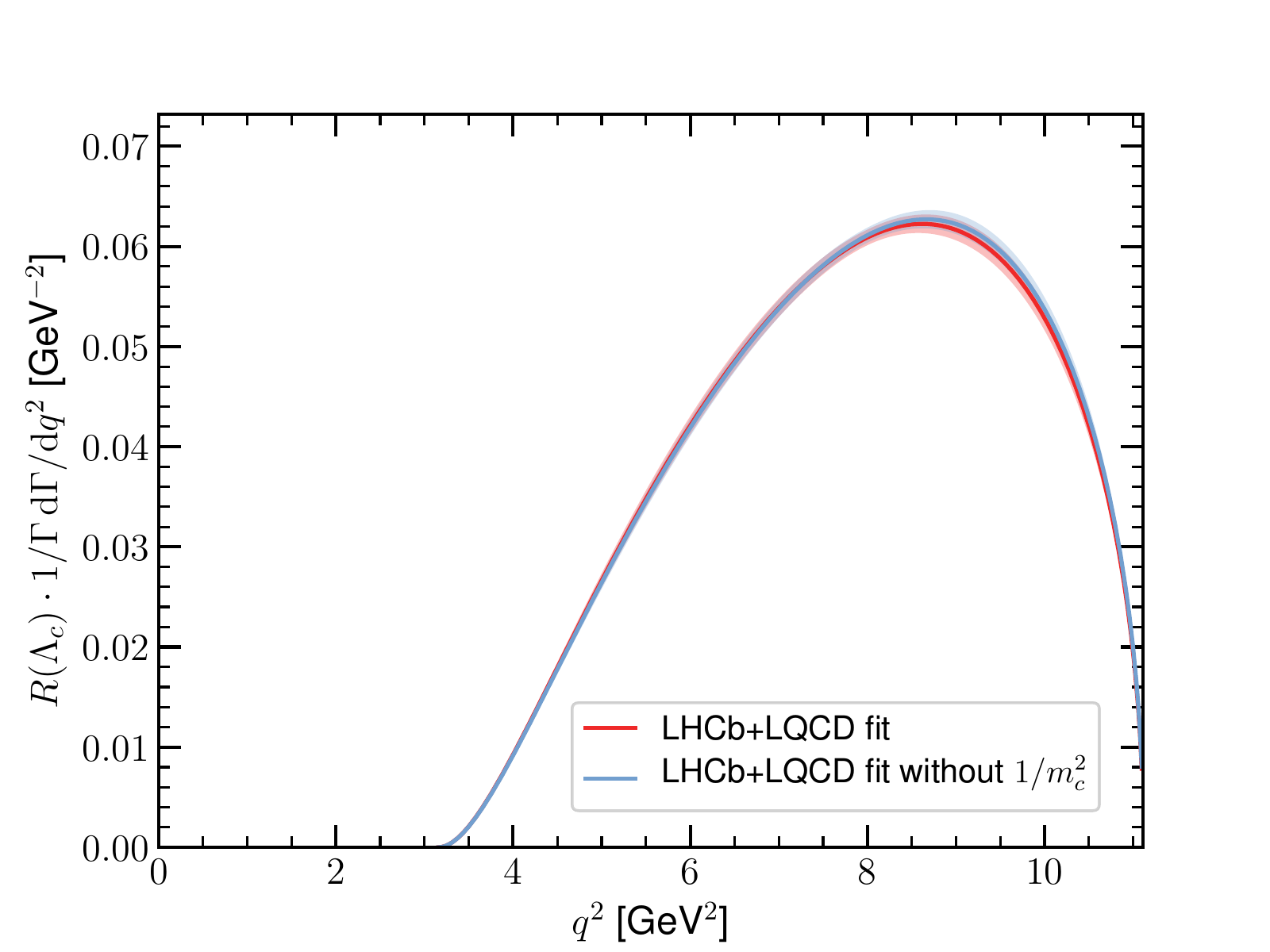}
\caption{Left: The data points show the LHCb measurement of the normalized
$\d\Gamma(\Lambda_b\to \Lambda_c \mu\bar\nu)/\d q^2$
spectrum~\cite{Aaij:2017svr}.  The red band shows our fit of the HQET
predictions to these data~\cite{Aaij:2017svr} and to the LQCD form
factors~\cite{Detmold:2015aaa}.  The blue curve shows the fit results, setting
the order $\lqcd^2/m_c^2$ terms to zero.  The gray band shows the LQCD
prediction.  Right: Our prediction for $\d\Gamma(\Lambda_b\to \Lambda_c
\tau\bar\nu)/\d q^2$ normalized to $R(\Lambda_c)$ from the same fit, with and
without including the $\lqcd^2/m_c^2$ terms.}
\label{fig:q2spec}
\end{figure}

\begin{figure*}[t]
\includegraphics[width=0.43\textwidth, clip, bb=0 0 420 315]{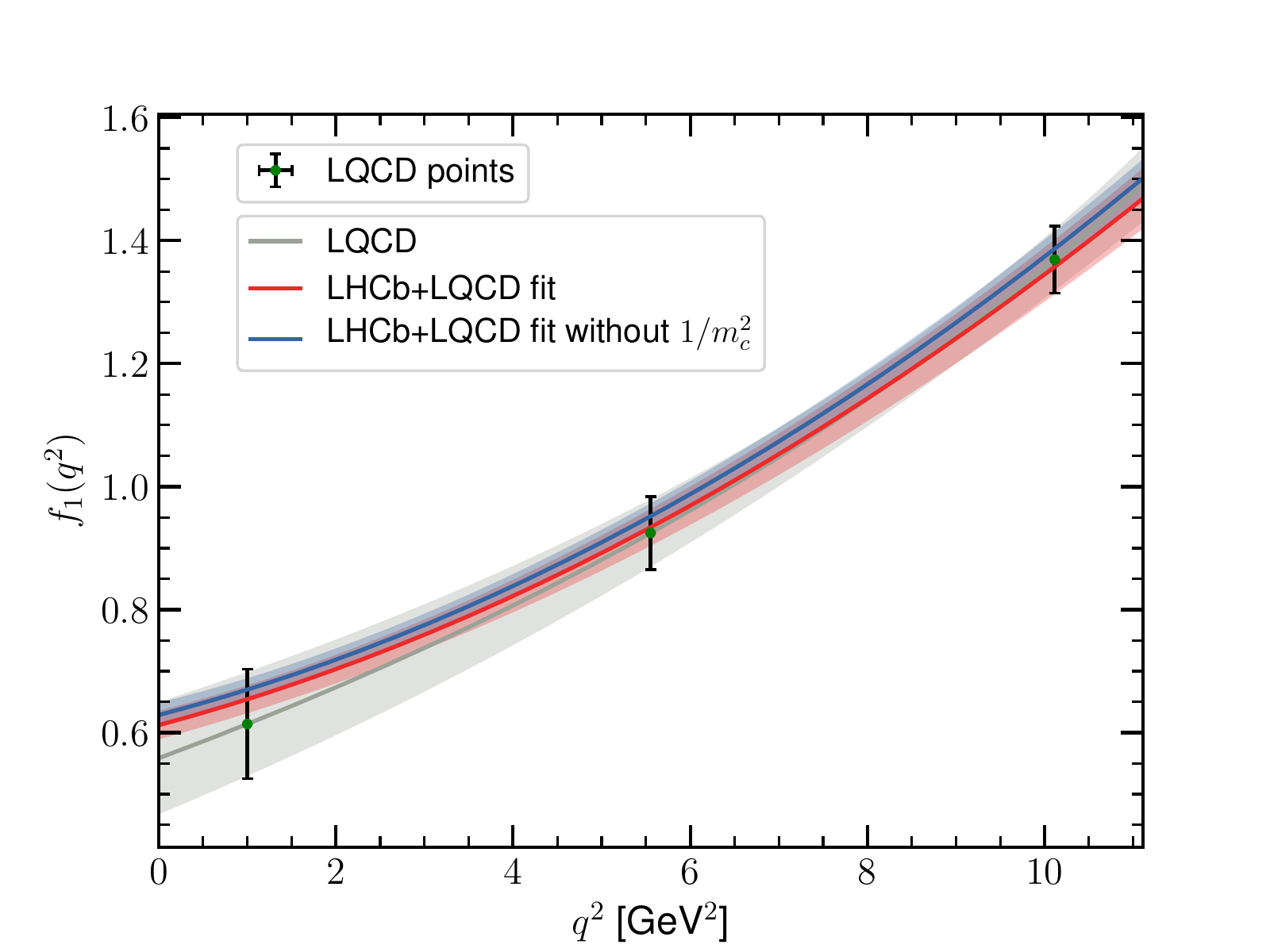}\hfil
\includegraphics[width=0.43\textwidth, clip, bb=0 0 420 315]{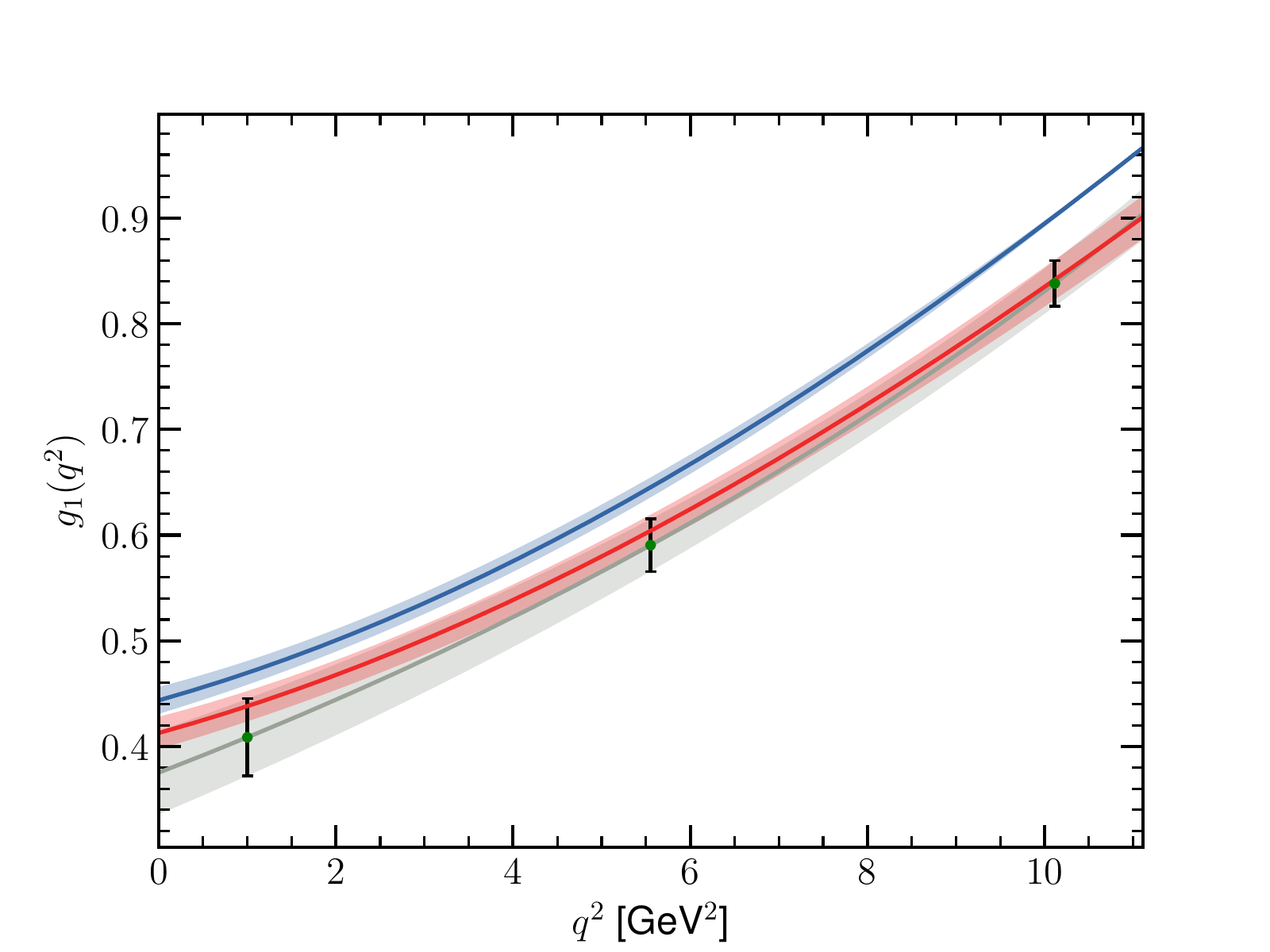}
\\[4pt]
\includegraphics[width=0.43\textwidth, clip, bb=0 0 420 315]{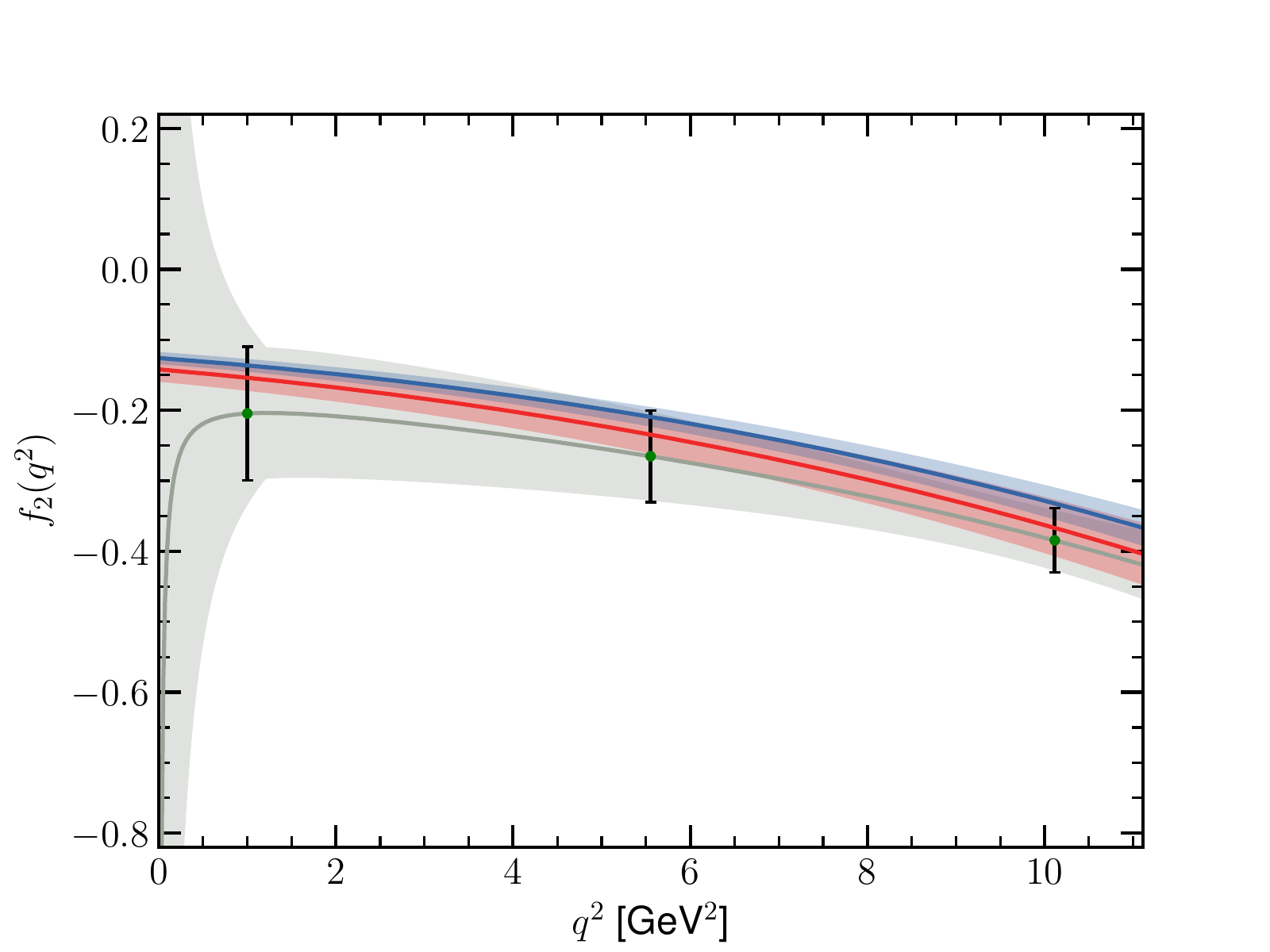}\hfil
\includegraphics[width=0.43\textwidth, clip, bb=0 0 420 315]{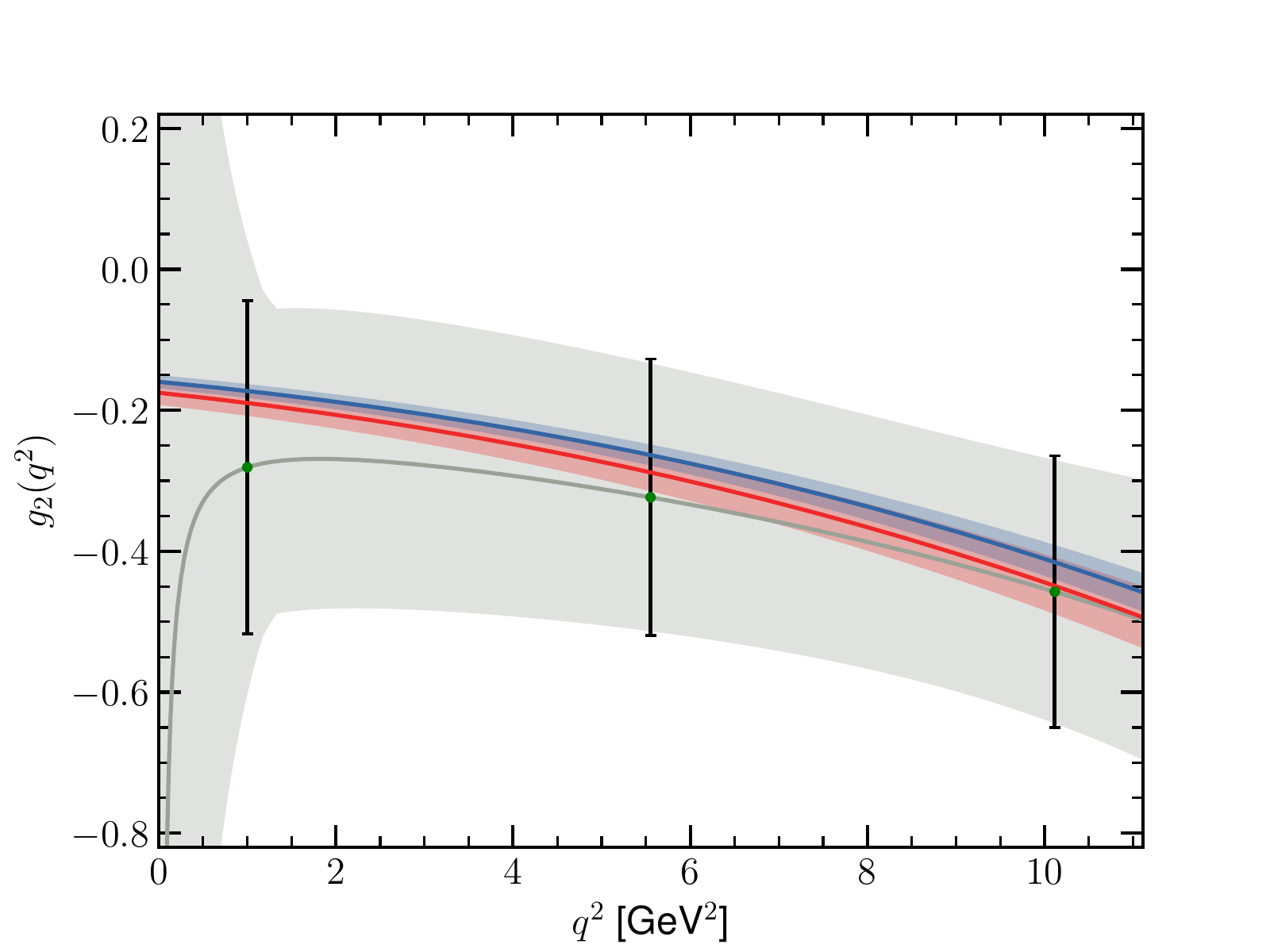}
\\[4pt]
\includegraphics[width=0.43\textwidth, clip, bb=0 0 420 315]{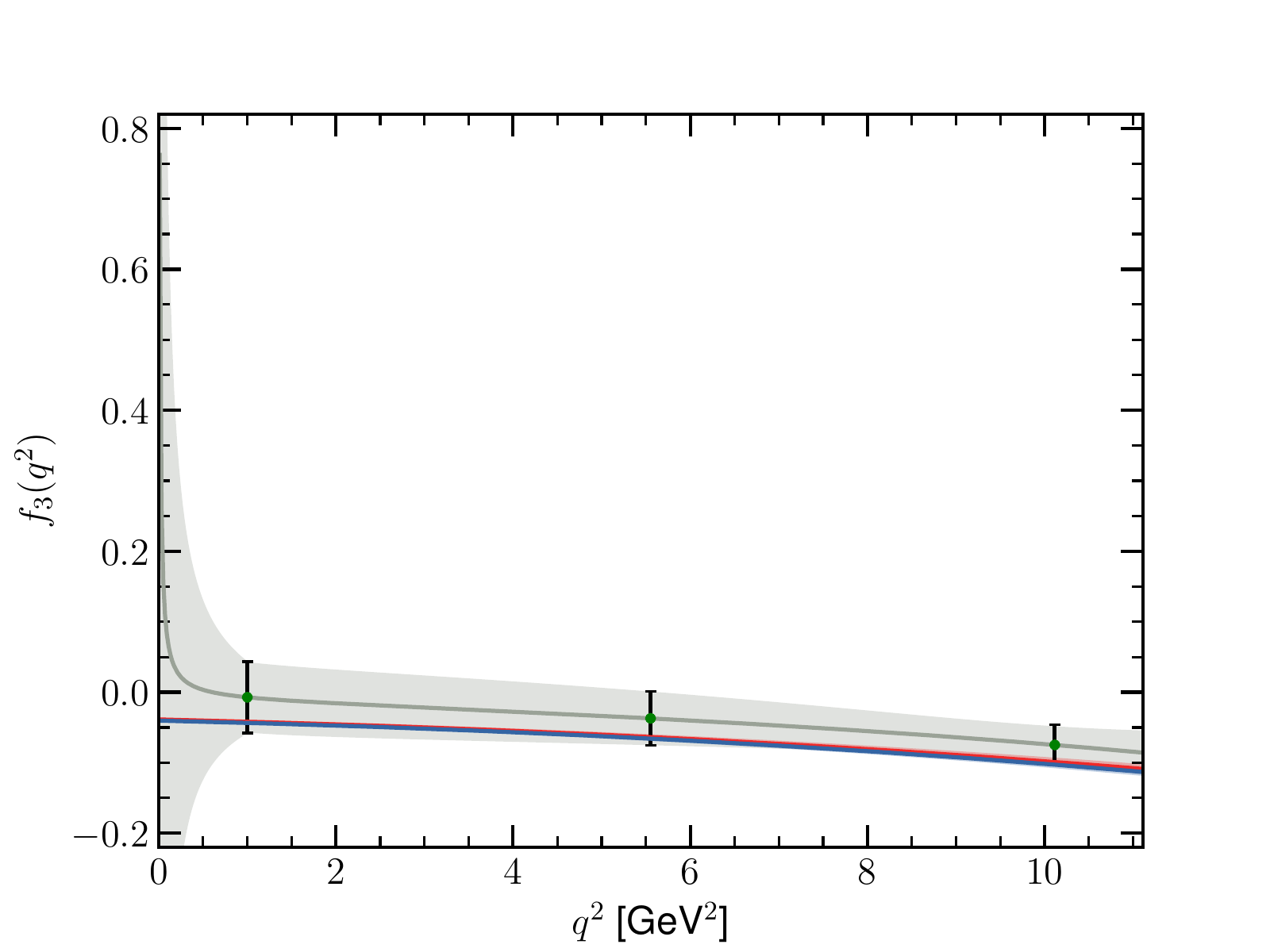}\hfil
\includegraphics[width=0.43\textwidth, clip, bb=0 0 420 315]{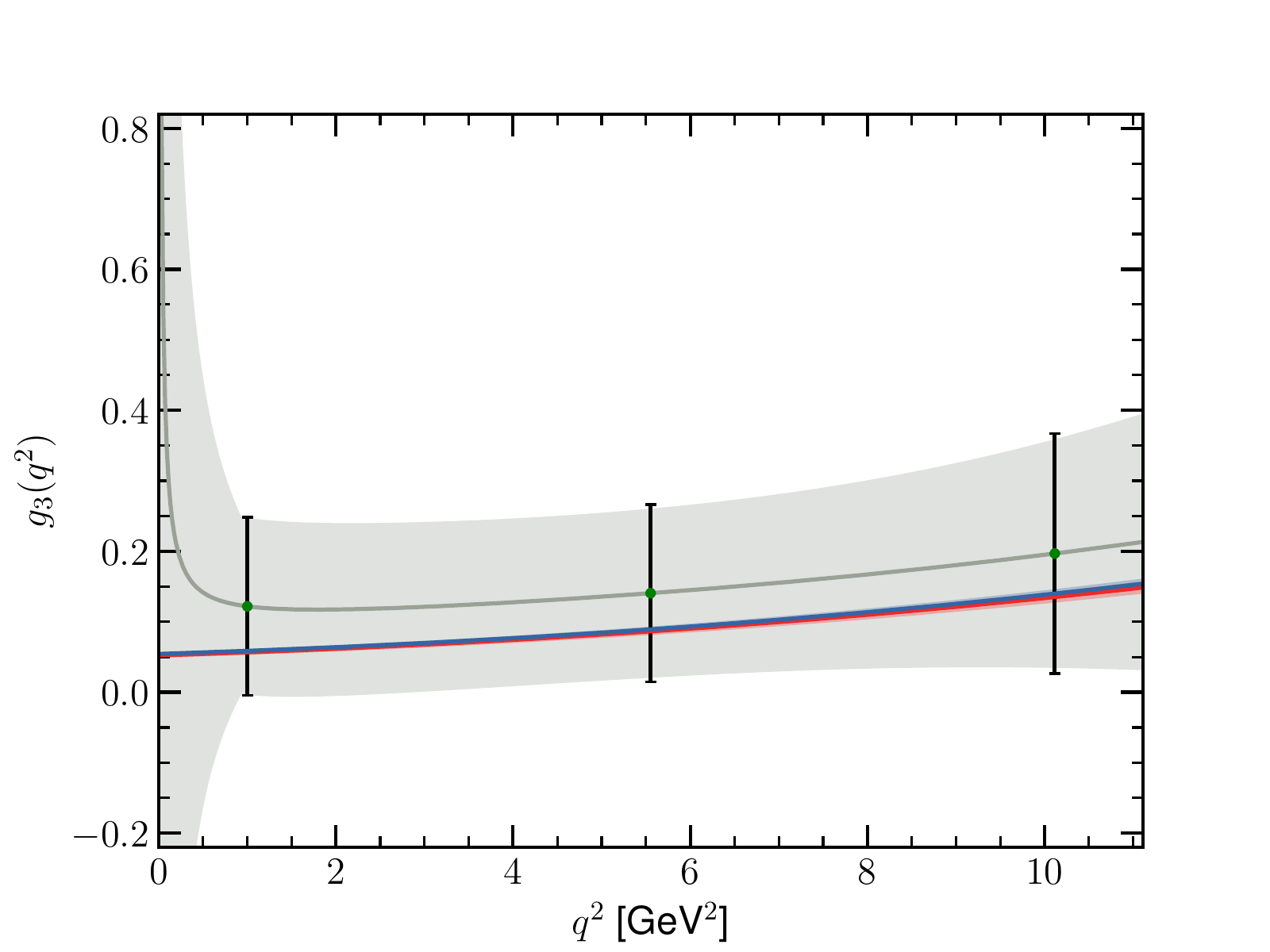}
\caption{Fits of the HQET predictions in Eq.~(\ref{ffexpsm}) to the LQCD
results~\cite{Detmold:2015aaa} for the 6 form factors (red bands) for
$f_{1,2,3}$ (left column) and $g_{1,2,3}$ (right column).  The blue bands show
the same fits, setting the order $\lqcd^2/m_c^2$ terms to zero.  Also shown are
the LQCD predictions (gray bands and data points); see text for details.}
\label{fig:lqcd_fits}
\end{figure*}

The methods used to fit $\d\Gamma(\Lambda_b\to \Lambda_c \mu\bar\nu)/\d q^2$
measured by LHCb~\cite{Aaij:2017svr} and lattice QCD (LQCD) calculation of the
(axial)vector form factors~\cite{Detmold:2015aaa} were described in
Ref.~\cite{Bernlochner:2018kxh}, and are only briefly recapitulated here.  LHCb
measured the $q^2$ spectrum in 7 bins, normalized to unity~\cite{Aaij:2017svr},
reducing the effective degrees of freedom in the spectrum from 7 to 6.  This
measurement is shown as the data points in the left plot in
Fig.~\ref{fig:q2spec}. Our fits to the LHCb data use the measured and predicted
partial rates in each bin. This procedure differs slightly from the fits
performed by LHCb~\cite{Aaij:2017svr}, which used the square root of $dN_{\rm
corr}/dw$ evaluated at the midpoint in the seven unfolded $w$ bins. The right
plot in Fig.~\ref{fig:q2spec} shows our prediction for $1/\Gamma \times
\d\Gamma(\Lambda_b\to \Lambda_c \tau\bar\nu)/\d q^2$, normalized to
$R(\Lambda_c)$.

The lattice QCD results~\cite{Detmold:2015aaa} for the six (axial)vector form
factors are published as fits to the BCL parametrization~\cite{Bourrely:2008za},
using either 11 or 17 parameters. We derive predictions for $f_{1,2,3}$ and
$g_{1,2,3}$ using the 17 parameter result at three $q^2$ values, $q^2 = \big\{ 1
\, \text{GeV}^2, \ q^2_{\rm max}/2, \ q^2_{\rm max} - 1 \, \text{GeV}^2\big\}$
for a total of eighteen form factor values,
constructing a covariance matrix from their correlation structure.   The values
of $q^2$ are chosen to sample both ends and the middle of the $q^2$ spectrum.
Adding more $q^2$ values from the BCL fit of the LQCD result to our sampling
does not noticeably affect the fit results. The difference in the form factor
values obtained using the 17 or the 11 BCL parameter results is added as an
uncorrelated uncertainty. This slightly differs from the prescription in
Ref.~\cite{Detmold:2015aaa}, which used the maximal differences of the form
factor values between the two parametrizations,  and cannot preserve the
correlation structure between the form factor values.  The 18 form factor values
used in our fits are shown as data points in Fig.~\ref{fig:lqcd_fits}. The LQCD
predictions, following the prescription of Ref.~\cite{Detmold:2015aaa}, are
shown as gray bands.  The uncertainties are in good agreement.  Similarly, the
gray band in Fig.~\ref{fig:q2spec} (left plot) shows the LQCD prediction for the
normalized spectrum, using the BCL parametrization.

In our fits, $m_b^{1S}$ and $\delta m_{bc}$ are constrained using Gaussian
uncertainties.  The leading order Isgur-Wise function is fitted to quadratic
order in $w-1$
\begin{equation}
\zeta = 1 + (w-1)\, \zeta' + \frac12 (w-1)^2\, \zeta''\,.
\end{equation} 
Alternative expansions using the conformal parameters $z$ or
$z^*$~\cite{Boyd:1995sq, Boyd:1997kz, Boyd:1997qw, Bourrely:2008za} instead of
$w$ yield nearly identical fits.  Therefore, we do not explore the differences
in the unitarity bounds between meson and baryon form
factors~\cite{Boyd:1995tg}.  Fits with $\zeta$ linear in either $w$, $z$, or
$z^*$ are poor, while adding more $q^2$ values to our sampling indicates no
preference for the inclusion of higher order terms in $w-1$. In the fits $\hat
b_{1,2}$ are assumed to be constants, which is appropriate at the current level
of sensitivity.  With better experimental and lattice constraints in the future,
the sensitivity to lifting these assumptions should be tested.

Fit results combining the LHCb and LQCD results are shown in
Table~\ref{tab:fitsummary}, and in Fig.~\ref{fig:lqcd_fits} by red bands. To
test the importance of the $\lqcd^2/m_c^2$ terms, we also perform a fit with the
order $\lqcd^2/m_c^2$ terms, parametrized by $\hat b_{1,2}$, set to zero. These
fits are shown in Fig.~\ref{fig:lqcd_fits} as blue bands, and the corresponding
fit values are provided in Table~\ref{tab:fitsummary}.  This is a much poorer
fit, changing $\chi^2/{\rm ndf}$ from $7.2/20$ to $18.8/22$.

\begin{table}[t]
\renewcommand{\arraystretch}{.75}
\newcolumntype{C}{ >{\centering\arraybackslash } m{4cm} <{}}
\begin{tabular}{c|CC}
\hline\hline
 & LHCb + LQCD & LHCb + LQCD\\
\hline 
$\zeta'$ & $ -2.04 \pm 0.08$ & $-2.06 \pm 0.08$	\\
$\zeta''$ &   $ 3.16 \pm 0.38$ & $3.28 \pm 0.36$ \\
$\hat b_1$/GeV${}^2$ & $ -0.46 \pm 0.15$ & $0^*$	\\
$\hat b_2$/GeV${}^2$ & $ -0.39 \pm 0.39$  & $0^*$ \\ \hline
$m_b^{1S}$/GeV & $ 4.72 \pm 0.05$ & $4.69 \pm 0.04$  \\
$\delta m_{bc}$/GeV & $ 3.40 \pm 0.02$  & $3.40 \pm 0.02$   \\
\hline
$\chi^2/\text{ndf}$ & $ 7.20 / 20$ & $18.8/22$ \\ 
\hline
$\RL$ &  $0.3237 \pm 0.0036$  & $0.3252 \pm 0.0035$\\
\hline\hline
\end{tabular}
\caption{HQET parameters extracted from the two fits discussed in the text. 
Predictions for $\RL$ for each fit are shown in the last row. The $\hat b_{1,2}$
values marked with an asterisk were fixed to zero in the fit; see text for
details.}
\label{tab:fitsummary}
\end{table}

We do not include explicitly an uncertainty for neglected higher order terms in
Eqs.~\eqref{ffexpsm} and \eqref{ffexpbsm}.  Four form factors, $f_3$, $g_3$,
$h_3$, and $h_4$ receive no $\lqcd^2/m_c^2$ corrections, so the agreement of
$f_3$ and $g_3$ with the LQCD results in the plots in the bottom row in
Fig.~\ref{fig:lqcd_fits} indicates that these higher order corrections are probably small. 
The order $\varepsilon_c\, \varepsilon_b$ corrections to $f_3$ and $g_3$ are
given by two new functions of $w$, $b_5$ and $b_6$~\cite{Falk:1992ws}, while the
$\varepsilon_c^3$ corrections to $f_3$ and $g_3$ also vanish.  Thus, including
such corrections, $b_5$ and $b_6$ would simply accommodate the $0.5\,\sigma -
1\,\sigma$ differences between the LQCD results and our fit for $f_3$ and
$g_3$.  The impact of this is small, for example, setting $f_3 = 0$
does not perceptibly change the SM prediction for $\RL$ compared to Eq.~(\ref{RLdef}), 
while setting $g_3 = 0$ changes the SM prediction from $\RL = 0.324 \pm
0.004$ in Eq.~\ref{RLdef} by about $1\sigma$, to $0.320\pm0.003$.

\begin{figure*}[t!]
\includegraphics[width=0.43\textwidth, clip, bb=0 0 420 315]{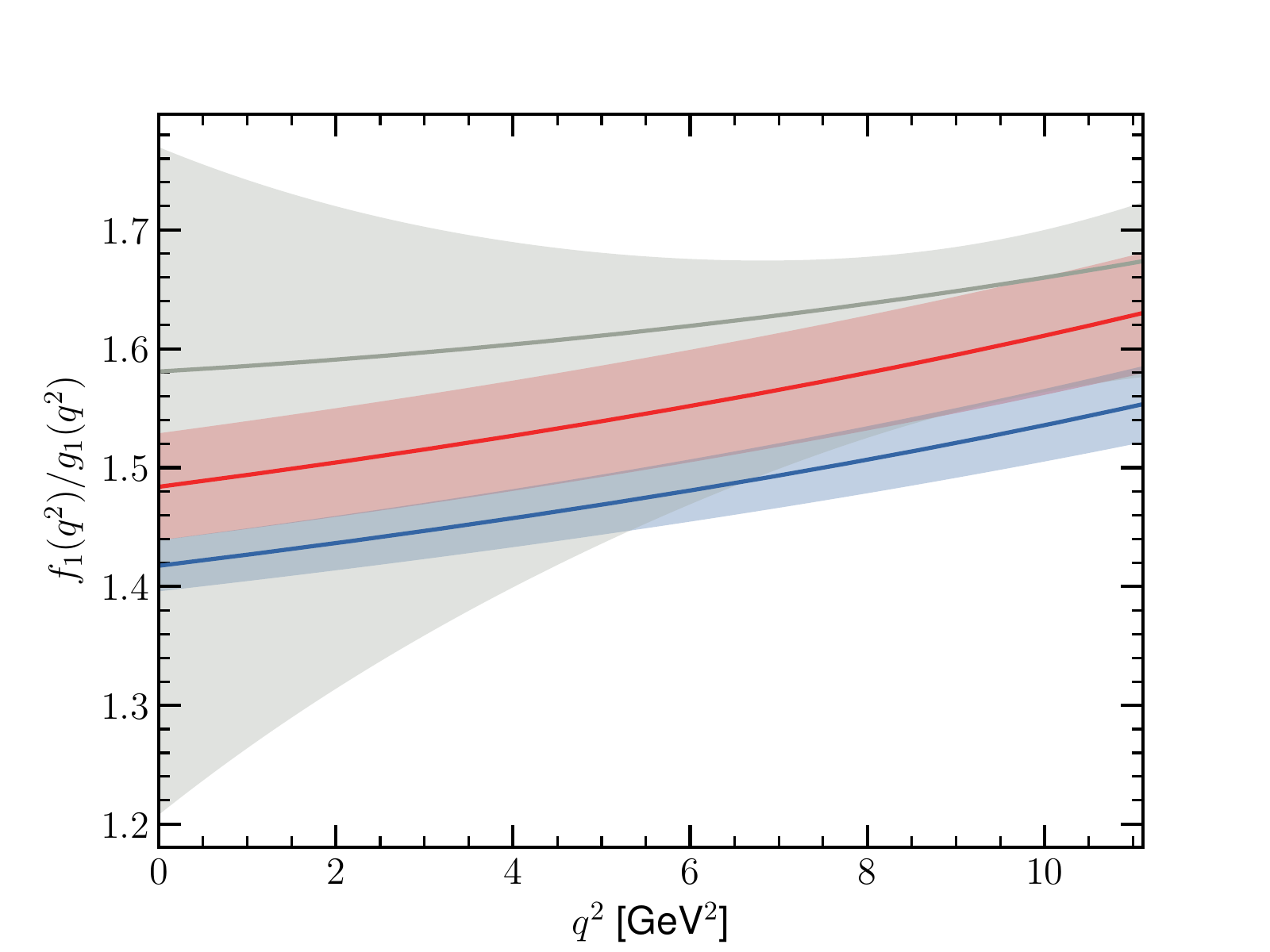}
\\[8pt]
\includegraphics[width=0.43\textwidth, clip, bb=0 0 420 315]{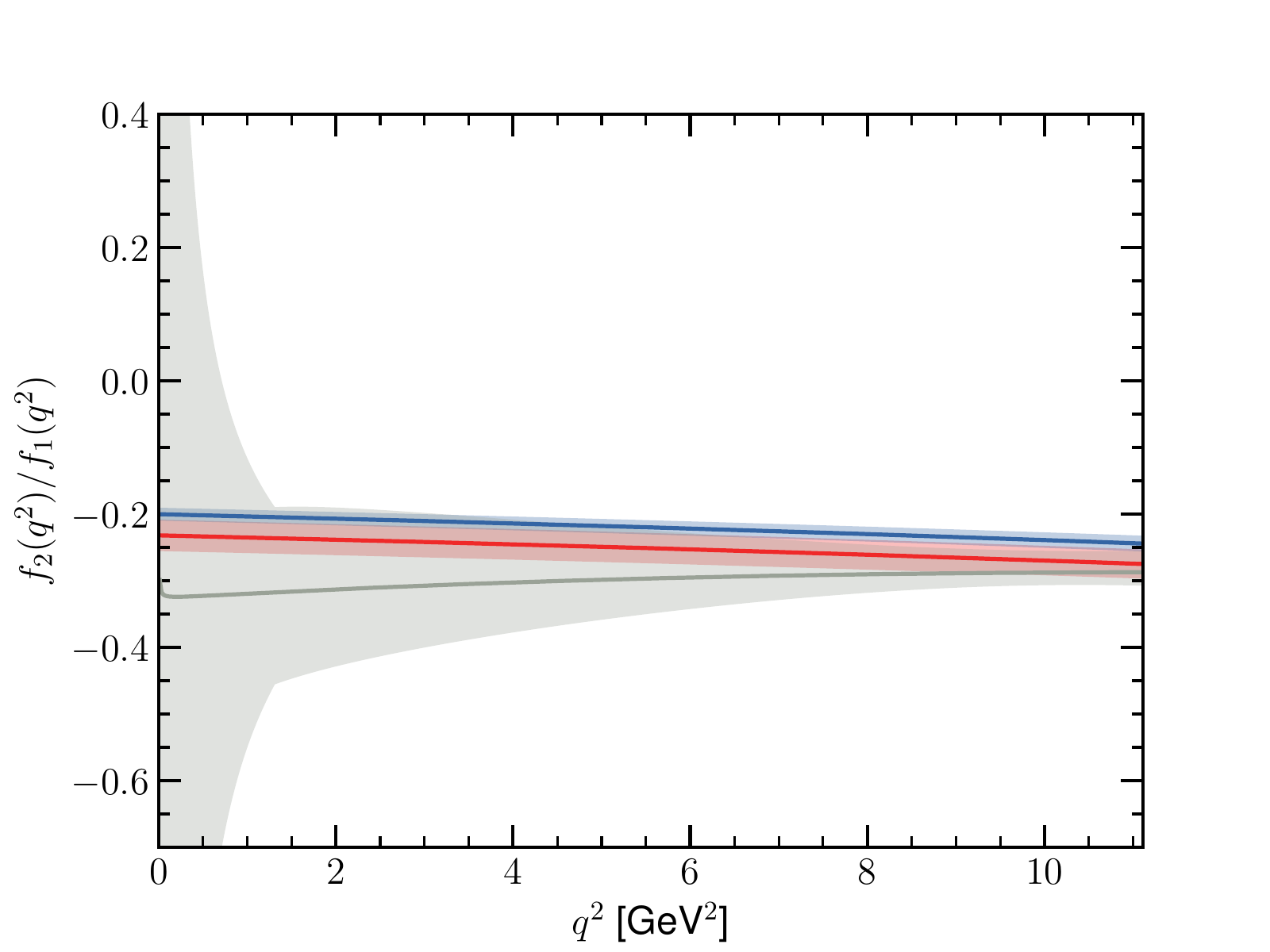}\hfil
\includegraphics[width=0.43\textwidth, clip, bb=0 0 420 315]{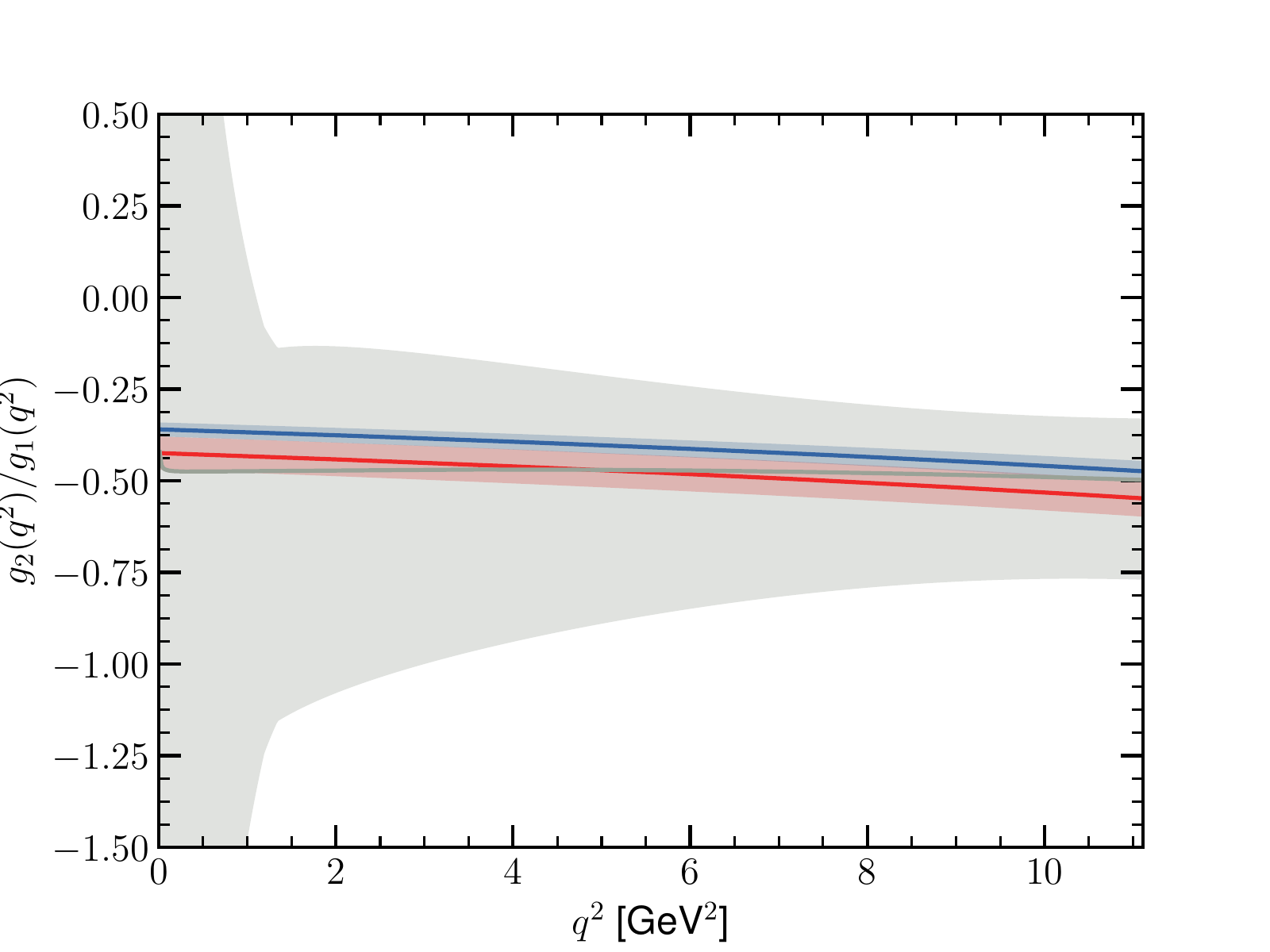}
\\[8pt]
\includegraphics[width=0.43\textwidth, clip, bb=0 0 420 315]{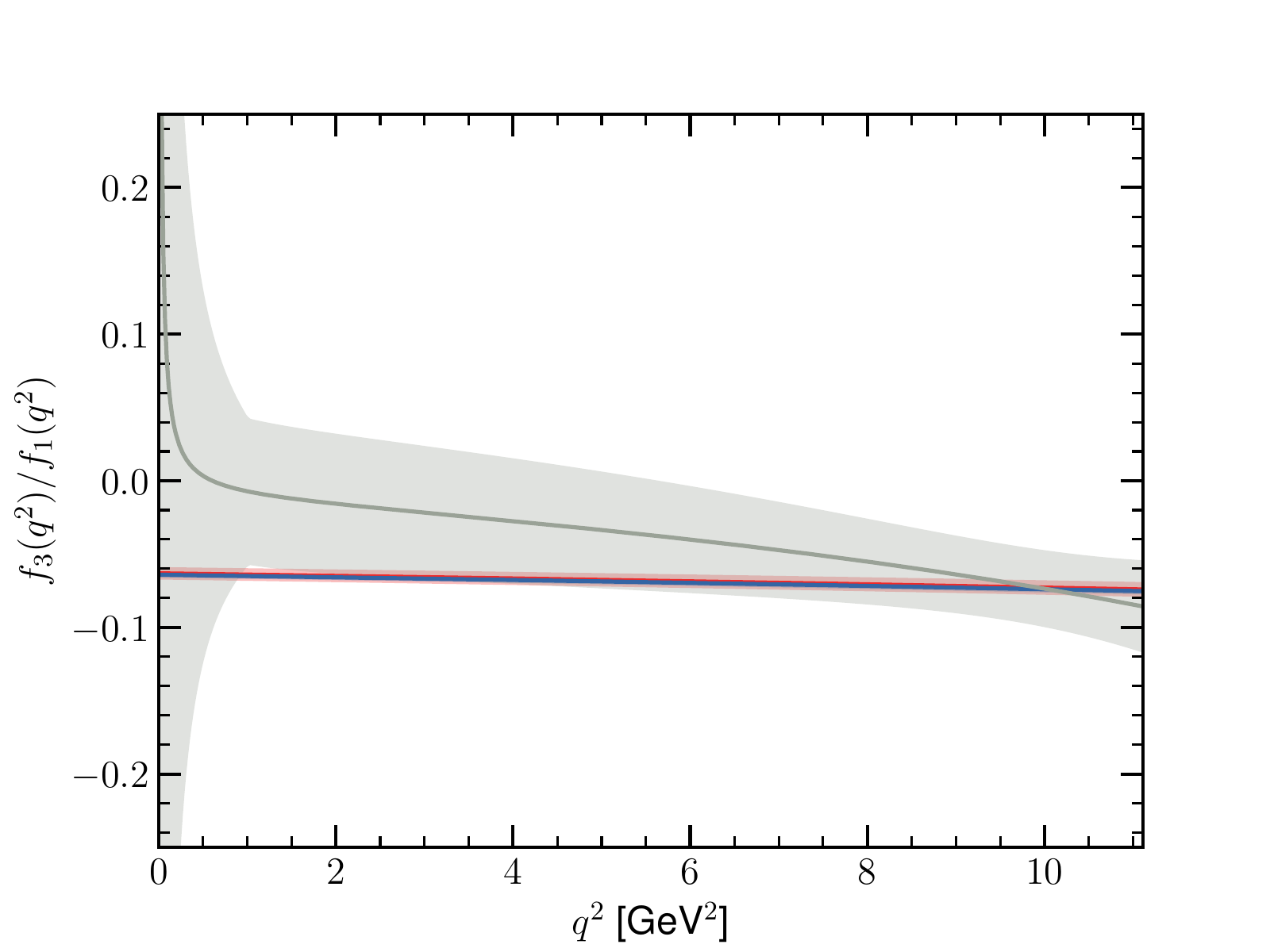}\hfil
\includegraphics[width=0.43\textwidth, clip, bb=0 0 420 315]{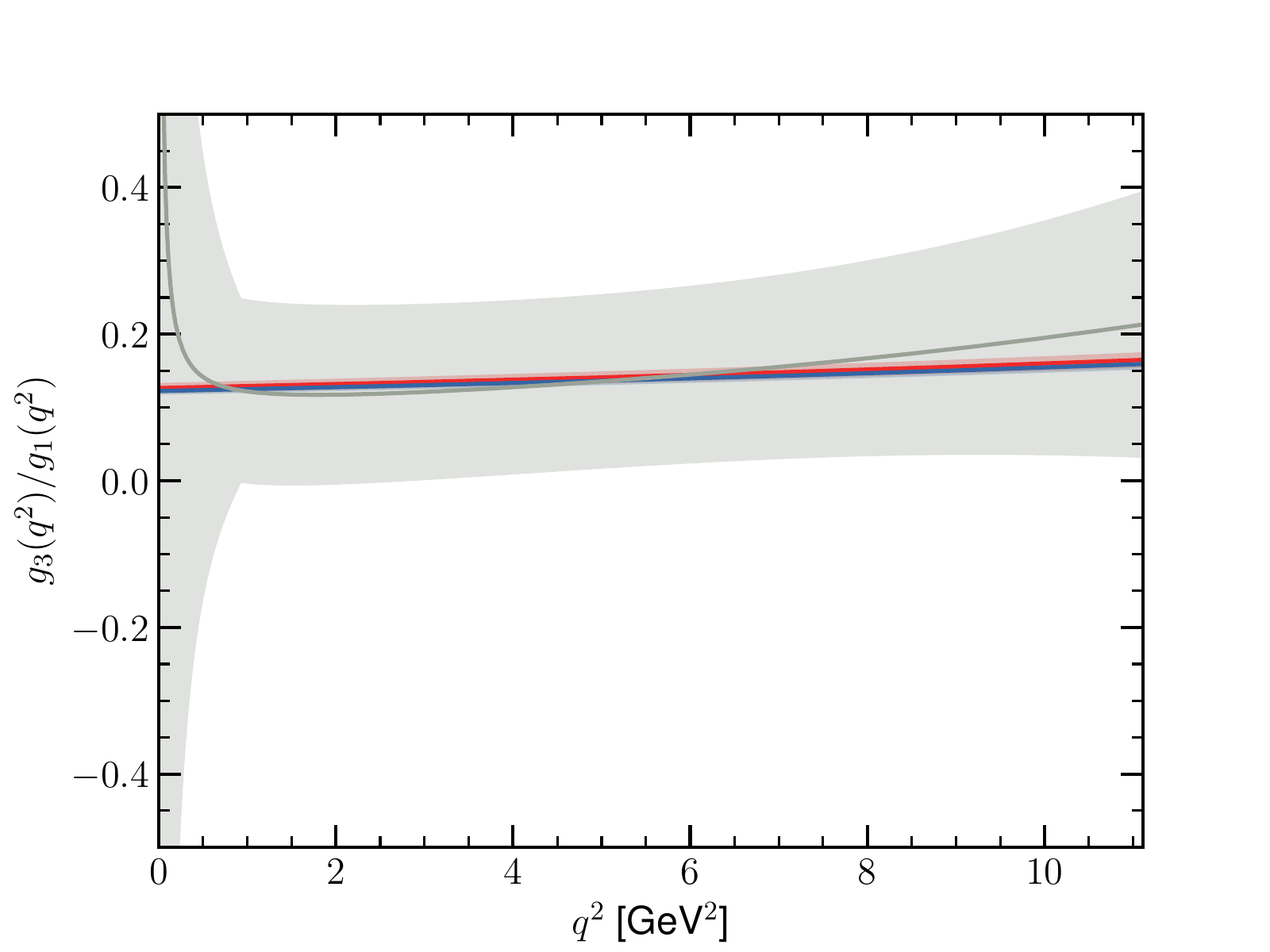}
\caption{Fits of the HQET predictions in Eq.~(\ref{ffexpsm}) to the LQCD
results~\cite{Detmold:2015aaa}, for five ratios of the six form factors.  The
top row shows $f_1/g_1$, which is ${\cal O}(1)$ in HQET, whereas $f_{2,3}/f_1$
(left column) and $g_{2,3}/g_1$ (right column) are expected to be ${\cal O}(\aS,
\lqcd/m_Q)$.  The red bands show our nominal fit including $\lqcd^2/m_c^2$
terms; the blue bands show fit results with $\lqcd^2/m_c^2$ terms set to zero.}
\label{fig:ffratios}
\end{figure*}

In Fig.~\ref{fig:ffratios} show our fit results for ratios of form factors
(red bands) and the LQCD predictions (gray bands).  The top plot shows  $f_1 /
g_1$, which HQET predicts to be ${\cal O}(1)$, whereas the four ratios $f_2 /
f_1$ and $g_2 / g_1$ (second row) and $f_3 / f_1$ and $g_3 / g_1$ (third row)
are predicted to be ${\cal O}(\varepsilon_{c,b},\, \alpha_s)$. 
The ratio, $f_1 / g_1\, (= f_\perp / g_\perp)$, is determined by
Eq.~(\ref{ffexpsm}) as
\beq\label{f1g1ratio}
\frac{f_1(w)}{g_1(w)} = 1 + \haS\, \big(C_{V_1} - C_{A_1}\big)
  + \big(\varepsilon_c + \varepsilon_b\big) \frac2{w+1} + \ldots\,,
\eeq
so the enhancement of $f_1$ relative to $g_1$ is a model independent prediction
of HQET, as seen in the top plot in Fig.~\ref{fig:ffratios}.

\subsection{Tensor form factors}

\begin{figure*}[t]
\includegraphics[width=0.43\textwidth, clip, bb=0 0 420 315]{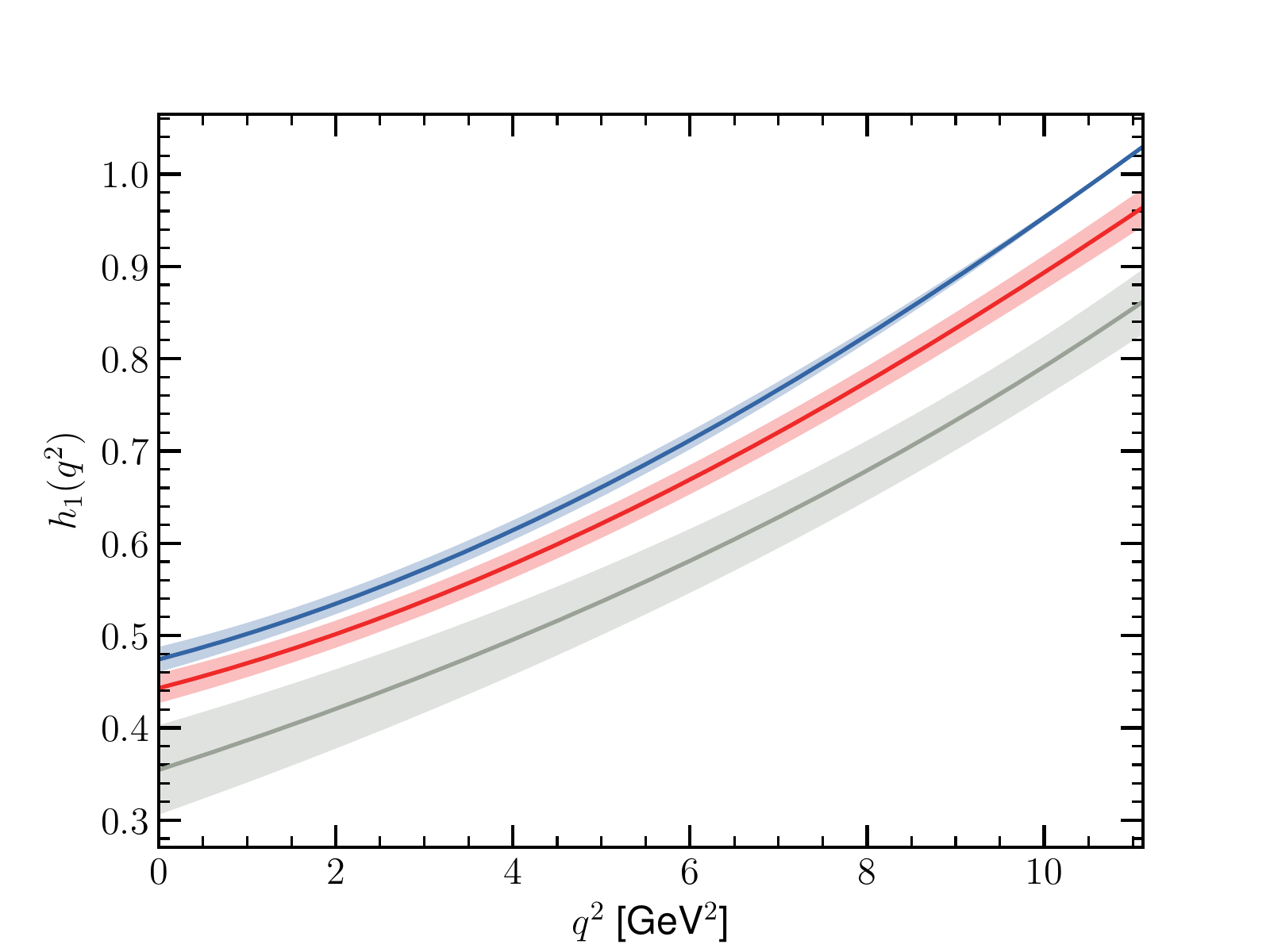}\hfil
\includegraphics[width=0.43\textwidth, clip, bb=0 0 420 315]{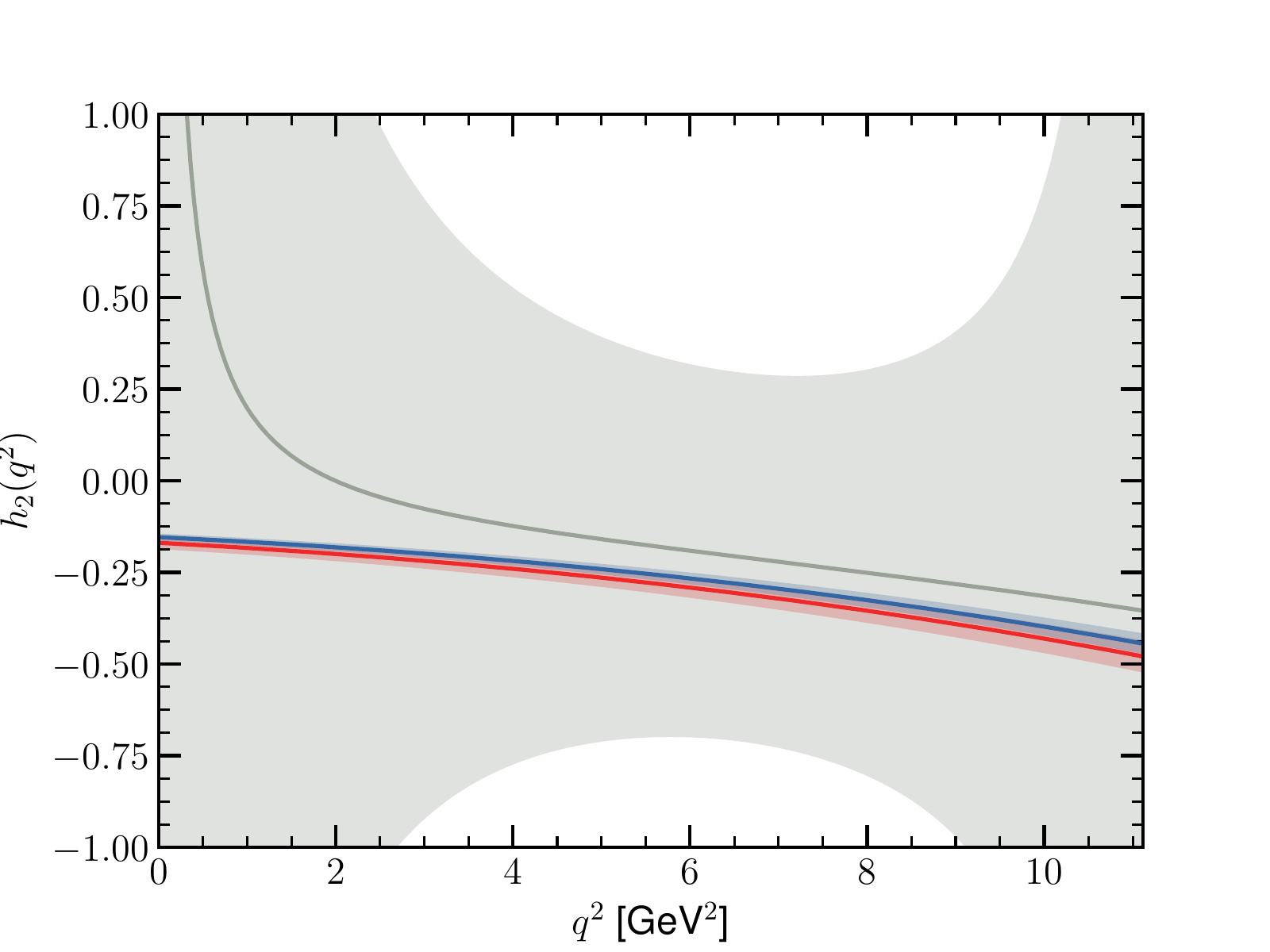}
\\[8pt]
\includegraphics[width=0.43\textwidth, clip, bb=0 0 420 315]{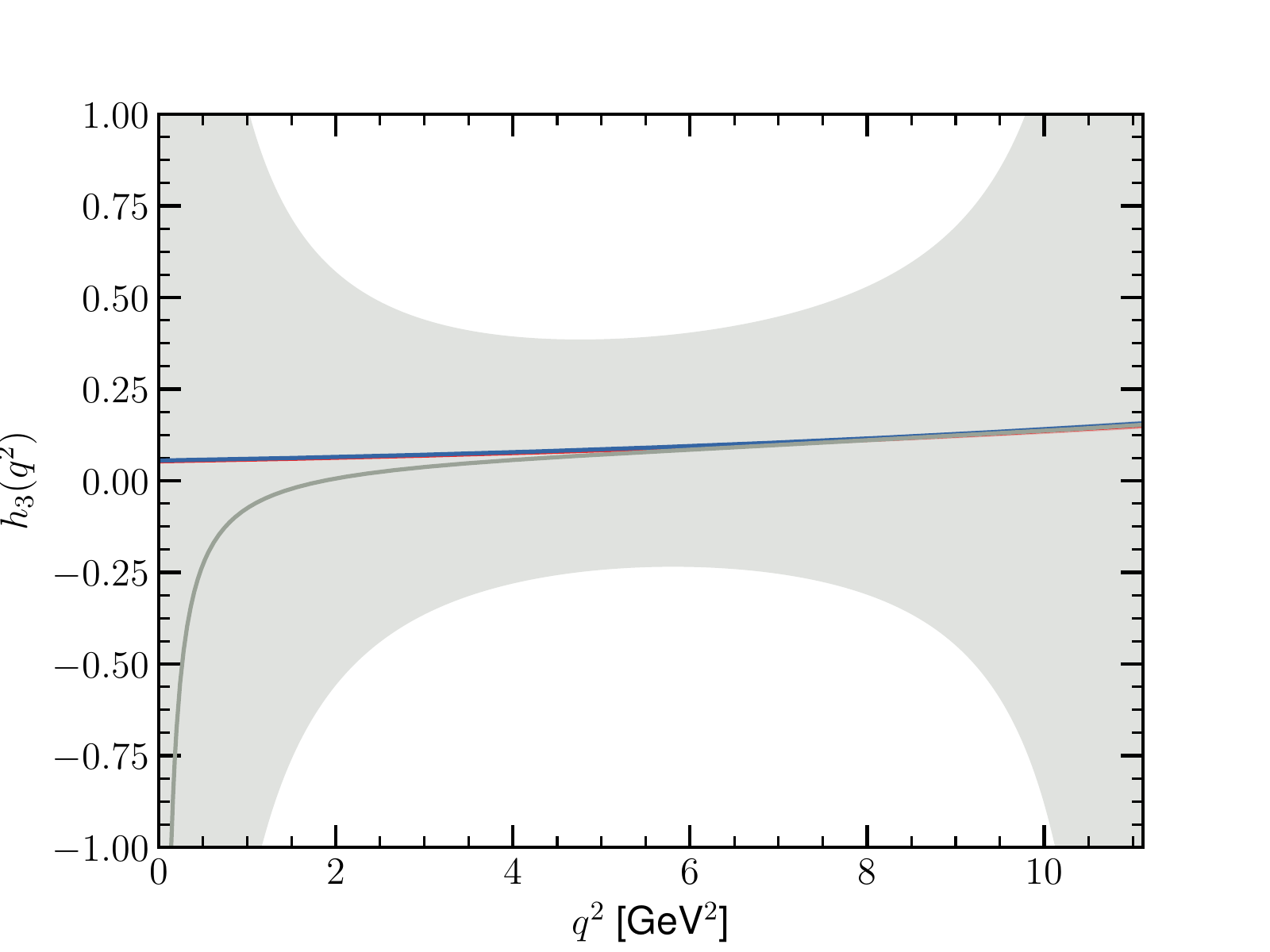}\hfil
\includegraphics[width=0.43\textwidth, clip, bb=0 0 420 315]{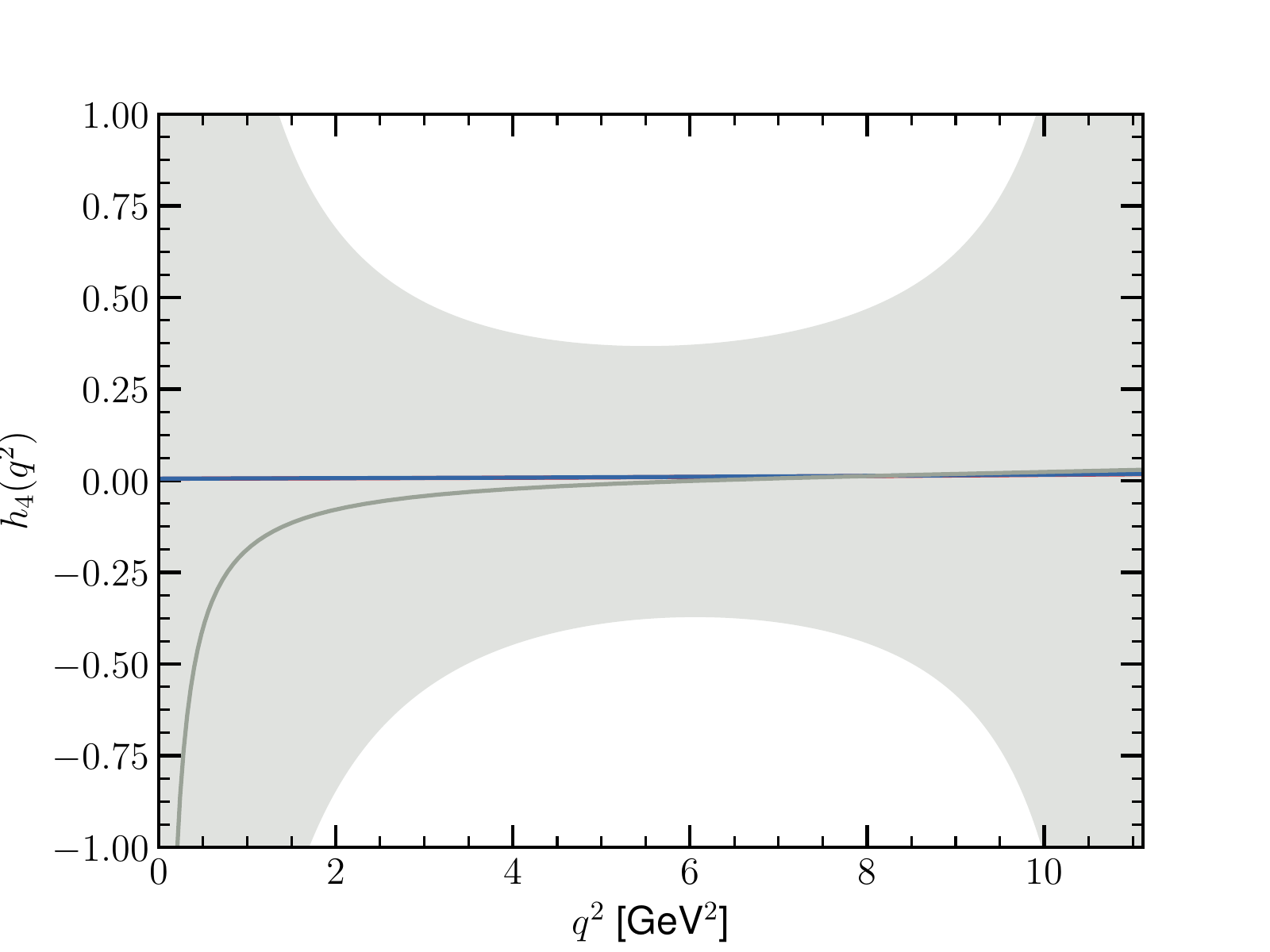}
\caption{Predictions for the tensor form factors based on Eq.~(\ref{ffexpbsm})
and our fit to the LHCb data and the LQCD calculation of the (axial)vector form
factors, overlayed with the LQCD calculation of the tensor form
factors~\cite{Datta:2017aue} (scaled to $\mu = \sqrt{m_b m_c}$).  The notation
is the same as in Fig.~\ref{fig:ffratios}.}
\label{fig:tensorff}
\end{figure*}

LQCD results~\cite{Datta:2017aue} for the tensor form factors are available, and
may be compared to  HQET predictions from our fits to the (axial)vector form
factors, via Eqs.~\eqref{ffexpbsm}.\footnote{ In Ref.~\cite{Datta:2017aue} the
equations of motions were used to express the scalar and pseudoscalar current
matrix elements in terms of the axial and vector currents. The resulting
expressions depend on the quark masses, $m_{b,c}$.  It is inconsistent beyond
leading order in $\alpha_s$ to use in such expressions the $\ov{\text{MS}}$
masses $\ov m_b(\ov m_b)$ and $\ov m_c(\ov m_c)$~\cite{Datta:2017aue} to
evaluate the decay rates. Instead, one must use $\ov m_c(\mu)$ and $\ov
m_b(\mu)$ at the same $\mu$.} The correspondence between the four form factors
used in this paper for the tensor current, $\big\{ h_1,\, h_2,\, h_3,\, h_4
\big\}$, defined in Eq.~(\ref{HQETffdefnew}), and those used in the LQCD
calculation~\cite{Datta:2017aue}, $\big\{ h_+,\, h_\perp,\, \widetilde h_+,\,
\widetilde h_\perp \big\}$, are given in Appendix~\ref{sec:FFapp}. In the former
basis, only one form factor, $h_1$, is nonzero in the heavy quark limit, while
the four form factors of the LQCD basis are equal to one another in this limit.
Note in particular that $h_1 = \widetilde h_+$.  

The LQCD results~\cite{Datta:2017aue} are presented using the BCL
parametrization, including the correlations of the parameters.  These results
are computed at the scale $\mu = m_b$, while in this paper we match HQET onto
QCD at $\mu = \sqrt{m_c m_b}$.  Since the tensor current has a nonzero anomalous
dimension, we use the multiplicative renormalization factor $\big[ \alpha_s(m_b)
/ \alpha_s(\sqrt{m_b m_c}) \big]^{4/25} \simeq 0.97$~\cite{Dorsner:2013tla,
Freytsis:2015qca}, in order to scale the form factors to $\mu = \sqrt{m_b\,
m_c}$.  

In Fig.~\ref{fig:tensorff} the gray bands show the LQCD results for the tensor
form factors converted to the $h_{1,2,3,4}$ basis. Our prediction from the fit
to the (axial)vector SM form factors and the LHCb data are overlaid as red
bands.  The LQCD uncertainties are large for $h_{2,3,4}$ at both ends of the
spectrum. This is an artifact of the $1/(w-1)$ and $1/q^2$ factors in the
transformation from the LQCD basis in Eq.~\eqref{tensorFFrelB}.  (The same
information in the $\big\{ h_+,\, h_\perp,\, \widetilde h_+,\, \widetilde
h_\perp \big\}$ basis is shown in Fig.~\ref{fig:tensorffback} in
Appendix~\ref{sec:FFapp}.  In this basis the uncertainties are not strongly
$q^2$ dependent.) Unlike the fits in Sec.~\ref{sec:fitLHCbVA}, the LQCD results
for the tensor form factors are not an input to our fits, so there is no free
parameter in these comparisons.  Figure~\ref{fig:tensorffback} shows that the
order $\varepsilon_c$ terms, which are fully determined by HQET in Eq.~(\ref{ffexpbsm}),
combined with the definitions in Eq.~(\ref{tensorFFrel}), account for the near
equality of $\widetilde h_\perp$ and $\widetilde h_+$, the slight enhancement of
$h_\perp$, and the substantial enhancement of $h_+$.  The top left plot in
Fig.~\ref{fig:tensorff} shows a tension between our fit and the LQCD
determination of $h_1 = \widetilde h_+$, visible in all plots in
Fig.~\ref{fig:tensorffback}.  In addition, the LQCD result for $h_1$ prefers a
slightly smaller curvature than our prediction. This is similar to what is seen
for $f_1$ and $g_1$ in the top row of Fig.~\ref{fig:lqcd_fits}: The LQCD results
prefer a smaller curvature at small $q^2$. This is related to the observation
that LQCD rate in Fig.~\ref{fig:q2spec} falls more quickly at small $q^2$ than
the LHCb measurement.

\subsection{$\RL$ predictions with new physics}

\begin{figure}[t]
\includegraphics[width=8cm]{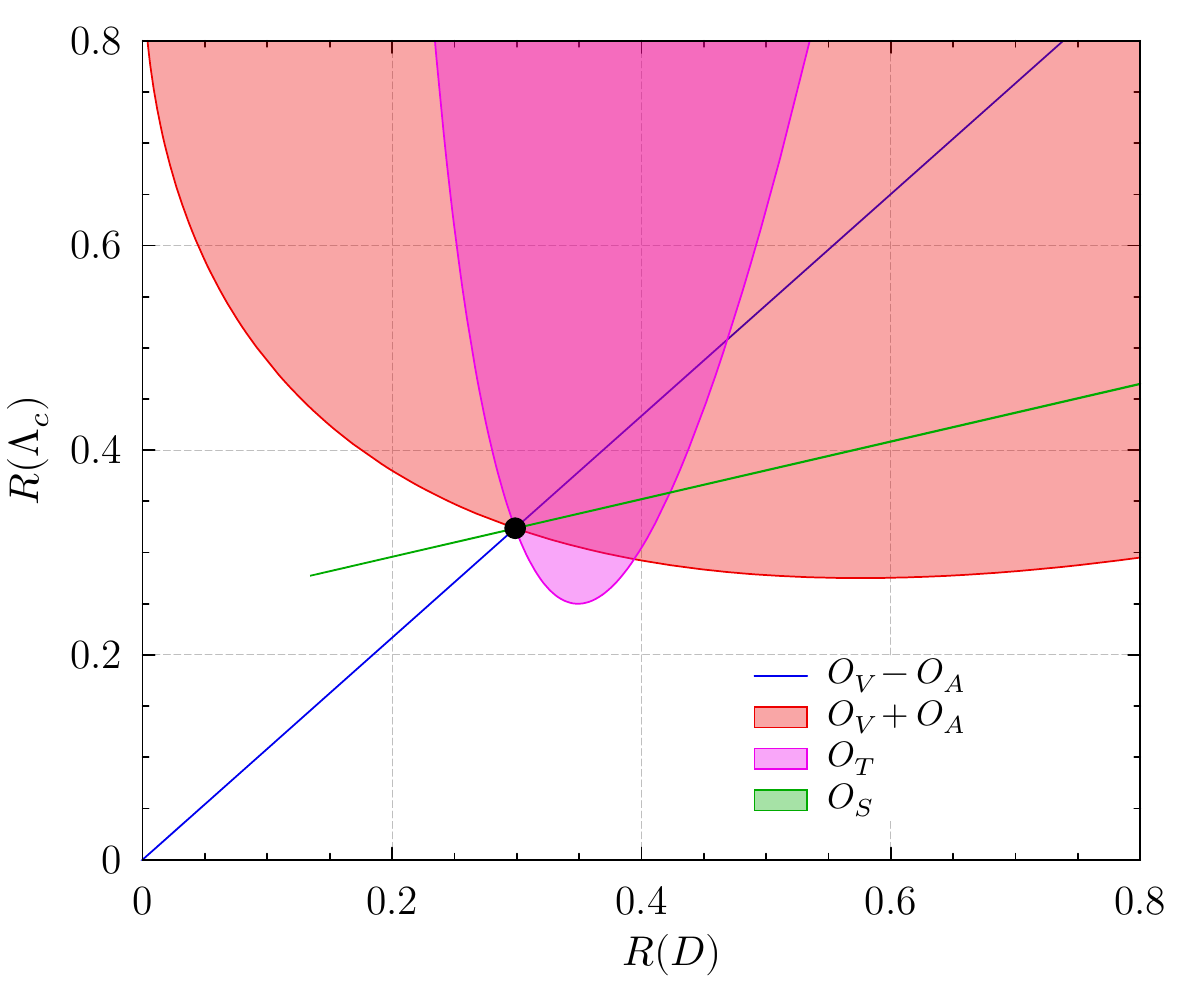}\hfill
\includegraphics[width=8cm]{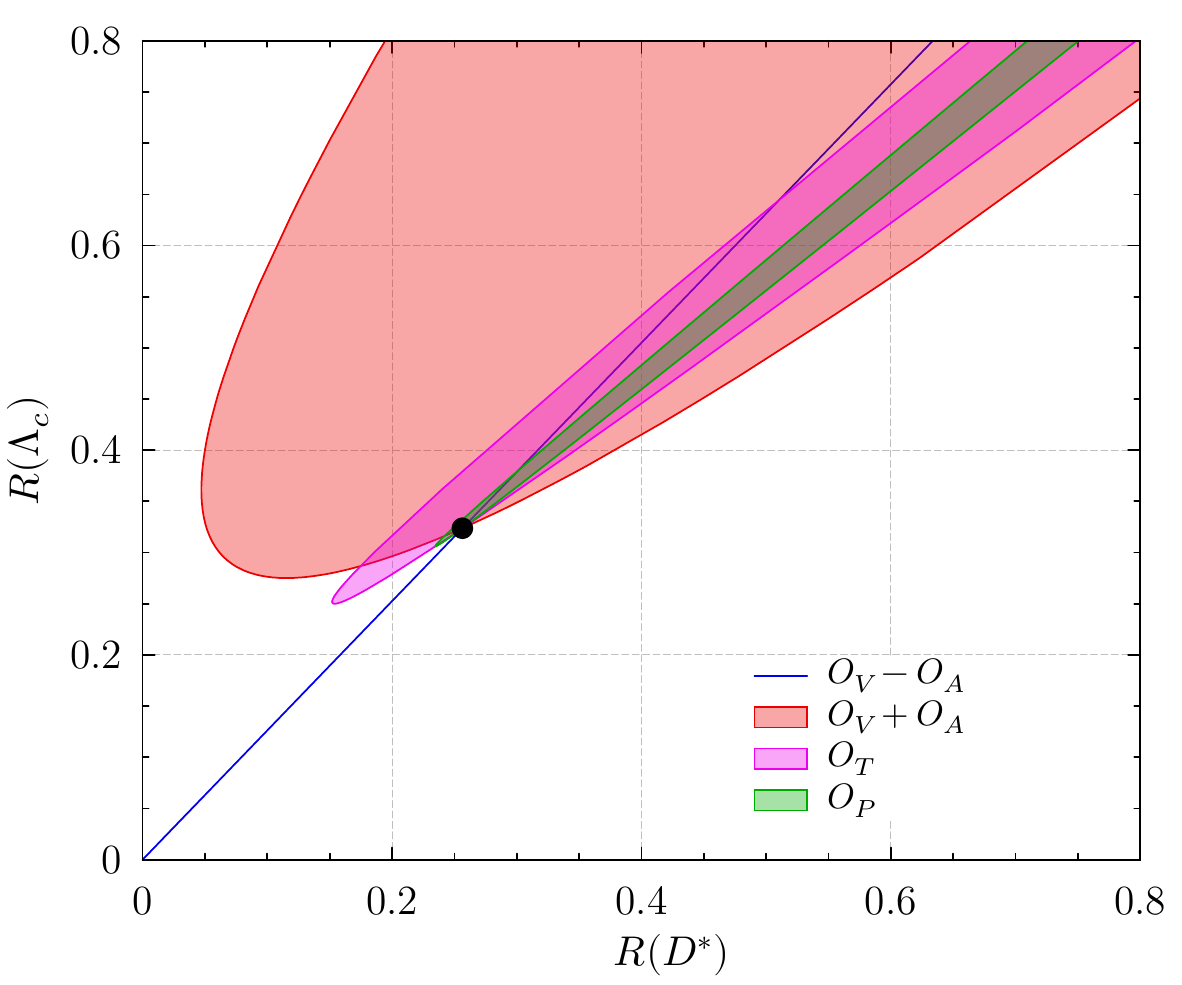}
\caption{$\RL$ vs.\ $R(D)$ (left) and $R(D^*)$ (right) for various NP
operators, in the basis defined in Eq.~\eqref{eqn:Odef}. The (pseudo)scalar
contributions vanish for the $D(D^*)$ modes, and are not shown.}
\label{fig:NPR}
\end{figure}

LHCb expects that the precision of the measurement of $\RL$ can compete with
that of $R(D^{(*)})$ in the future~\cite{LHCbprojection}.  For the SM prediction
we obtained~\cite{Bernlochner:2018kxh}
\begin{equation}
 \RL = 0.324 \pm 0.004 \,.
\end{equation}
Our form factor fit, combined with the expressions for the NP rates in Appendix~\ref{app:NPampl} 
and the HQET predictions in Eqs.~\eqref{ffexpbsm}, allows for precision computation of $\RL$ for
arbitrary NP contributions (see e.g. Refs~\cite{Shivashankara:2015cta,Dutta:2015ueb,Li:2016pdv} for prior analyses). 
To gain a sense of the sensitivity of $\RL$, in
Fig.~\ref{fig:NPR}, we show the allowed regions in the $\RL - R(D)$ and $\RL -
R(D^*)$ planes, as any one of the five NP couplings in Eq.~(\ref{eqn:Odef}) are
turned on.  The boundary of each region corresponds to real NP Wilson
coefficients, while the interior requires a relative phase between the SM and
NP.  The $V-A$ NP interaction cannot have a physical phase relative to the SM,
and therefore spans a line in the $\RL - R(D^{(*)})$ planes.  Possibly by
numerical coincidence,  the scalar operator exhibits a very large correlation
between $\RL$ and $R(D)$, resulting in a very narrow $\RL - R(D)$ region for
this operator. Note that the (pseudo)scalar contributions vanish for the $D\,
(D^*)$ modes, respectively, and are not shown.

\begin{figure}[t]
\includegraphics[width=8cm]{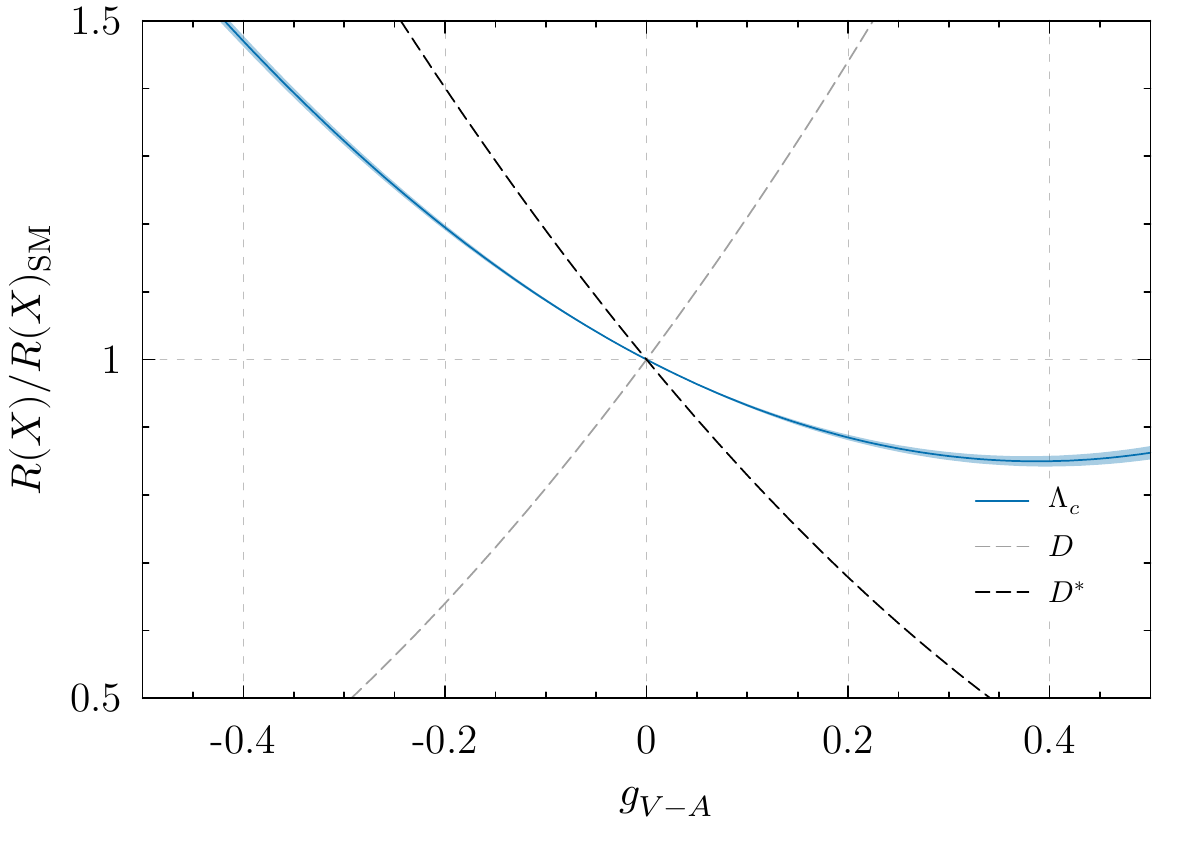}\hfill
\includegraphics[width=8cm]{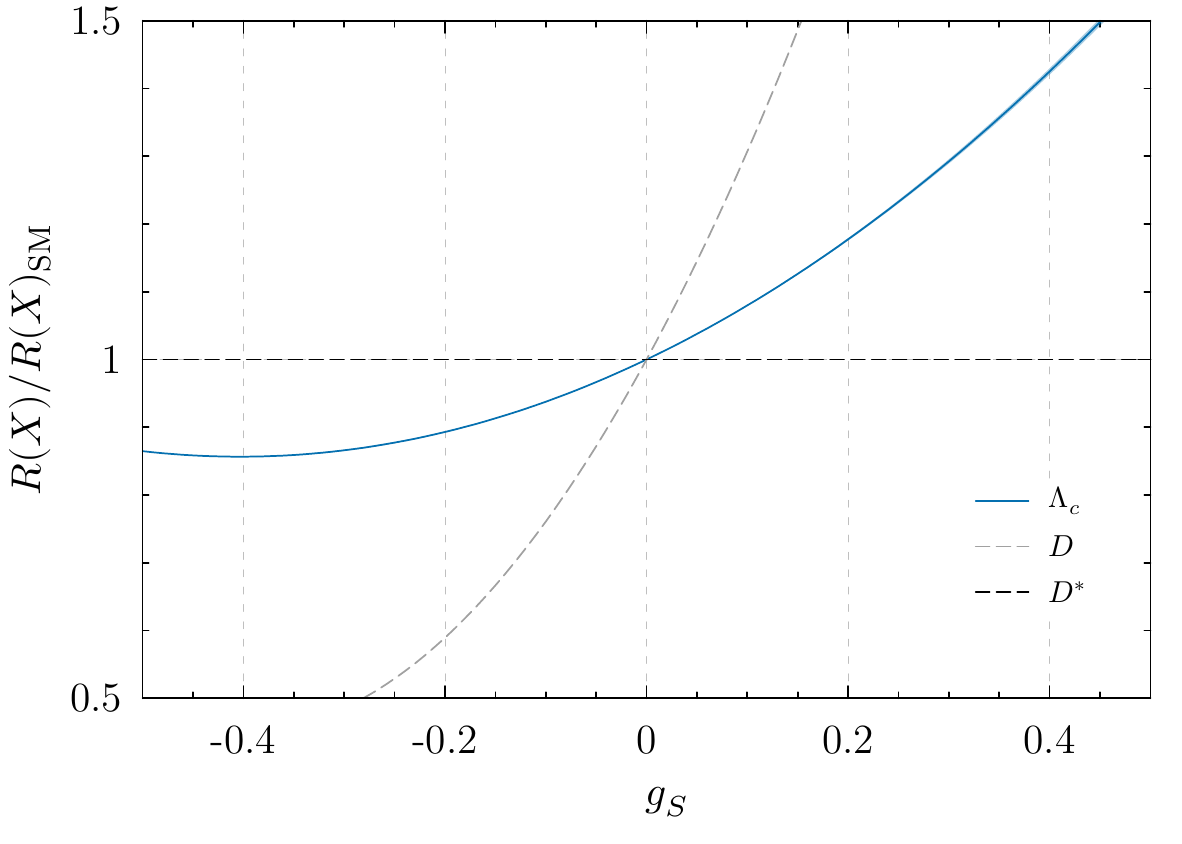}\\
\includegraphics[width=8cm]{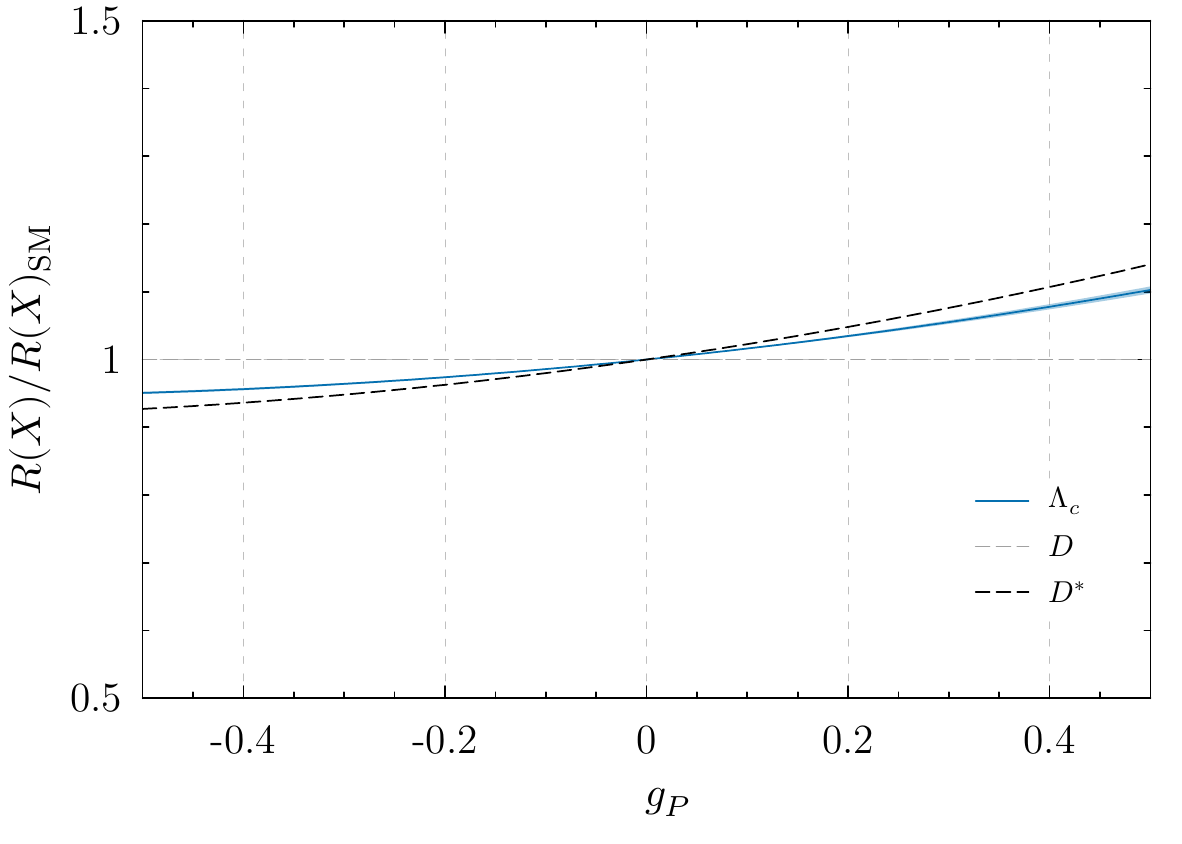}\hfill
\includegraphics[width=8cm]{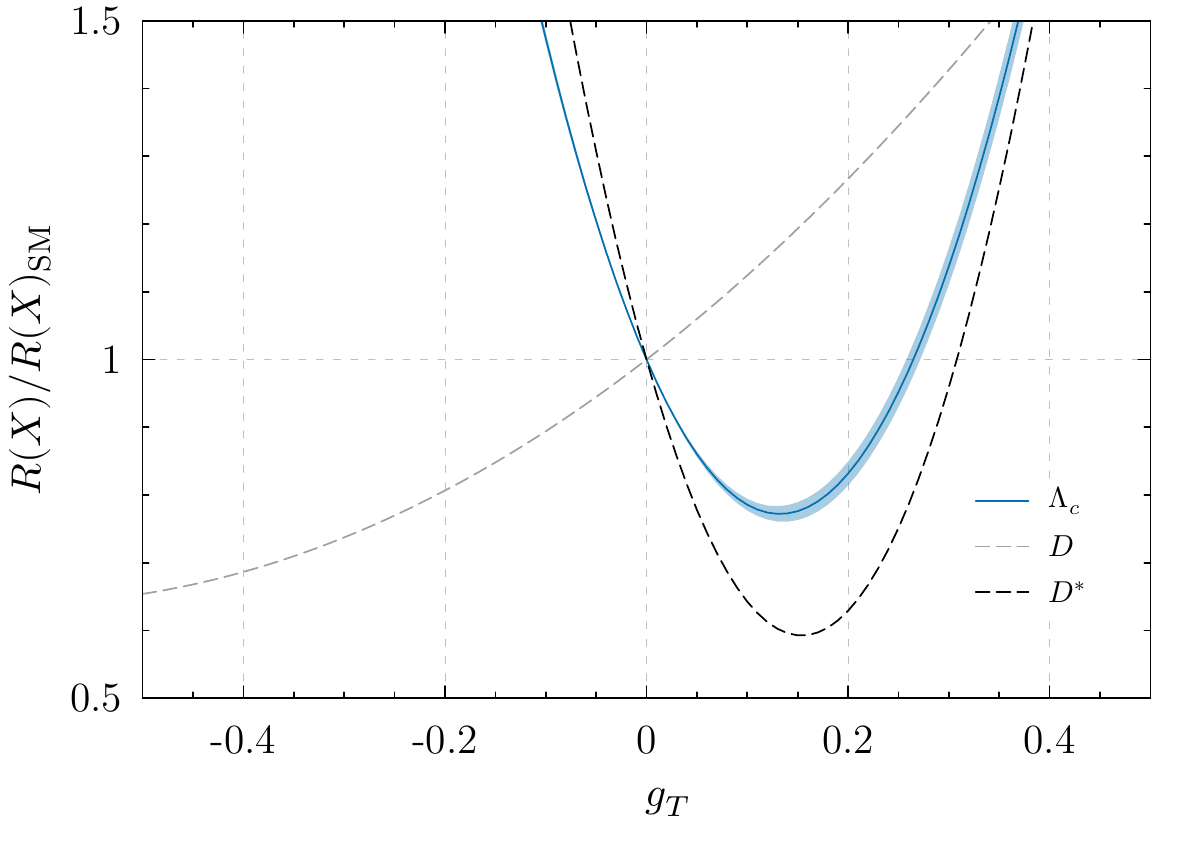}
\caption{$\RL/\RL_{\text{SM}}$ and $R(D^{(*)})/R(D^{(*)})_{\text{SM}}$
predictions for real NP couplings, in the operator basis of
Eq.~\eqref{eqn:Odef}.}
\label{fig:NPg}
\end{figure}

In Fig.~\ref{fig:NPg} we compare the variation in $\RL/\RL_{\text{SM}}$ with 
the corresponding ratios for $D^{(*)}$, as a function of each NP
coupling, assuming they are real.  An error band, corresponding to the uncertainties in the fit of
Ref.~\cite{Bernlochner:2018kxh}, is also shown.  In some cases the errors are
imperceptible.  We see that the NP sensitivity of $\RL$ is typically between
the $R(D^*)$ and $R(D)$ variations.  

\section{Factorization and $\Lambda_b\to\Lambda_c\pi$}

The LHCb measurement of the $\d\Gamma(\Lambda_b\to \Lambda_c^+ \mu^- \bar\nu)/\d
q^2$ spectrum~\cite{Aaij:2017svr} is normalized to unity, and the LCQD results
for the $\Lambda_b \to \Lambda_c$ form factors are also independent of
$|V_{cb}|$.  Thus, our fit is sensitive to hadronic parameters, but it cannot be
combined with the present LHCb data to extract $|V_{cb}|$.  One may, however,
use the LHCb measurement of $\d\Gamma(\Lambda_b\to \Lambda_c^+ \mu^- \bar\nu)/\d
q^2$ to test factorization in $\Lambda_b\to\Lambda_c\pi$, or to extract
$|V_{cb}|$ assuming factorization (see also Ref.~\cite{Leibovich:2003tw}).  For
$B \to D^{(*)}\pi$ decays, it has long been known that the ratios ${\cal B}(B^-
\to D^0\pi^-)/{\cal B}(\bar{B}^0 \to D^+\pi^-) \simeq 1.9$ and ${\cal B}(B^- \to
D^{*0}\pi^-)/{\cal B}(\bar{B}^0 \to D^{*+}\pi^-) \simeq 1.8$~\cite{PDG} deviate
substantially from unity, the prediction in the heavy quark limit.  This implies
that $\mathcal{O}(\lqcd/m_c)$ contributions to the amplitudes enter at the
$30\%$ level, and deviations from factorization in the heavy quark limit are
substantial.

At leading order in the heavy quark expansion, the $\Lambda_b \to \Lambda_c \pi$
matrix element factorizes such that the nonleptonic rate is related to the
semileptonic rate at $q^2 = m_\pi^2$ via
\begin{equation}
\label{Lbfactor2}
\Gamma(\Lambda_b\to \Lambda_c\pi) = 6\pi^2\, \big(C_1+C_2/3\big)^2\, 
  |V_{ud}|^2\, f_\pi^2\,
  \frac{\d\Gamma(\Lambda_b\to \Lambda_c e \bar\nu)}{\d q^2} 
  \bigg|_{q^2=m_\pi^2} \,, 
\end{equation}
where $f_{\pi}=131\,$MeV is the pion decay constant, and $C_{1,2}$ are the usual
Wilson coefficients in the effective Hamiltonian, satisfying $(C_1+C_2/3)\,
|V_{ud}| \simeq 1$.  (Uncertainties in this linear combination, $f_\pi$, and
$\tau_{\Lambda_b}$ are neglected.)  In Eq.~(\ref{Lbfactor2}), we write the
$\Lambda_c e \bar\nu$ final state to emphasize that the semileptonic rate has to
be evaluated neglecting lepton masses.  In $\Lambda_b\to \Lambda_c \mu \bar\nu$
decay, measured by LHCb, the impact of $m_\mu \neq 0$ is substantial at
$q^2=m_\pi^2$. 

\enlargethispage{-1\baselineskip}

Combining the factorization relation in Eq.~(\ref{Lbfactor2}), our fit for the
form factors, and $|V_{cb}| = \big(4.22 \pm 0.08 \big) \times
10^{-2}$~\cite{PDG} predicts ${\cal B}(\Lambda_b \to \Lambda_c\pi) = (3.6 \pm
0.3) \times 10^{-3}$, where this uncertainty is from the fit and $|V_{cb}|$. By
comparison, the measured nonleptonic branching ratio~\cite{PDG} is\footnote{This
PDG average for ${\cal B}(\Lambda_b \to \Lambda_c\pi)$ includes an uncertainty
scale factor of 1.5~\cite{PDG}, and is based on two LHCb~\cite{Aaij:2014lpa,
Aaij:2014jyk} and one CDF~\cite{Abulencia:2006df} measurements.  Reproducing
this is not easy, as it involves rescaling the CDF result from ${\cal
B}(\Lambda_c \to pK^-\pi^+) = (5.0 \pm 1.3)\%$ to the latest values: ${\cal
B}(\Lambda_c \to pK^-\pi^+) = (6.84 \pm
0.24^{+0.21}_{-0.27})\%$~\cite{Zupanc:2013iki} and ${\cal B}(\Lambda_c \to
pK^-\pi^+) = (5.87 \pm 0.27 \pm 0.23)\%$~\cite{Ablikim:2015flg}. The LHCb
measurements also preceded Ref.~\cite{Ablikim:2015flg}, and lifetime and other
data also changed.}
\begin{equation}\label{Lbnonlep}
{\cal B}(\Lambda_b \to \Lambda_c\pi) = (4.9 \pm 0.5) \times 10^{-3} \,.
\end{equation}
Conversely, assuming factorization, one could use Eqs.~\eqref{Lbnonlep} in
Eq.~\eqref{Lbfactor2} to extract $|V_{cb}| = (4.9\pm 0.3) \times 10^{-2}$,  where
this uncertainty is only from our form factor fit and the measured branching
fraction, without an uncertainty assigned to the factorization relation itself.
Thus we observe an $\mathcal{O}(15\text{--}20\%)$ deviation from the
factorization relation in Eq.~\eqref{Lbfactor2}, consistent with this deviation
arising from a $\lqcd/m_c$ suppressed
correction~\cite{Bauer:2001cu}.\,\footnote{Regarding the behavior of the heavy
quark expansion, the decay constants also satisfy the HQET scaling better than
was thought in the 1990s.  The $N_f=2+1+1$ FLAG~\cite{[][{, and updates at
\url{http://flag.unibe.ch/}}.]Aoki:2016frl} averages, $f_B = (186 \pm 4)\,\MeV$
and $f_D = (212 \pm 1.5)\,\MeV$, yield $f_B/f_D \simeq 0.88$, which is not
inconsistent with the leading order HQET relation~\cite{Politzer:1988wp,
Shifman:1986sm} $\sqrt{m_D/m_B}\, [\alpha_s(m_b)/\alpha_s(m_c)]^{-6/25} \simeq
0.68$, plus $\lqcd/m_{c,b}$ corrections.}

\section{Conclusions}
\label{sec:concl}

Fitting the LHCb measurement of the normalized $q^2$ spectrum for
$\Lambda_b\to\Lambda_c\mu\nu$ decay~\cite{Aaij:2017svr}, and the six
(axial)vector form factors calculated in lattice QCD~\cite{Detmold:2015aaa}, one
can test HQET relations and the applicability of power counting.  In
Ref.~\cite{Bernlochner:2018kxh} we found that the $\lqcd^2/m_c^2$ corrections
were constrained by the fit to be of the expected magnitude, without any signs
of enhancements or breakdown of the power counting at the $m_c$ scale, as is
sometimes claimed in the literature. Compared to the lattice QCD only
determination of the SM prediction of $\RL$, by fitting the LHCb measurement as
well, we further found that the uncertainty of the SM prediction  may be
substantially reduced, generating the most precise SM prediction for $\RL$ to
date, $\RL= 0.324 \pm 0.004$.

We expanded and generalized the results of Ref.~\cite{Bernlochner:2018kxh} in
several ways.  First, we calculated $\Lambda_b\to\Lambda_c$ semileptonic form
factors for all four-Fermi NP operators, including the ${\cal O}(\lqcd^2/m_c^2)$
corrections (as well as the corresponding helicity amplitudes for use in the
\texttt{Hammer} library~\cite{Duell:2016maj}).  Using our fit of the LHCb
measurement and the LQCD prediction for the six (axial)vector SM form factors,
we obtained parameter-free predictions for the four tensor form factors at
$\mathcal{O}(\lqcd^2/m_c^2)$.  We observed some tension between our results
based on HQET and those in Ref.~\cite{Datta:2017aue}, at a magnitude greater
than the $\lqcd^2/m_c^2$ corrections (see the top left figure for the $h_1 =
\widetilde h_+$ form factor in Fig.~\ref{fig:tensorff}).

The small uncertainties in our fit to the (axial)vector form factors, combined
with HQET predictions for the form factors at $\mathcal{O}(\lqcd^2/m_c^2)$
allowed us to derive precise predictions for $\RL$ for arbitrary NP. We studied
the NP impacts on $\RL$, including their correlations with $R(D^{(*)})$.  The NP
sensitivity of $\RL$ typically falls between those of $R(D^*)$ and $R(D)$.  We
also explored tests of factorization in $\Lambda_b \to \Lambda_c\pi$ decay. 
Factorization in the heavy quark limit, combined with $|V_{cb}|$ measurements and our fit to the
semileptonic form factors, implies a mildly lower nonleptonic rate than is measured, 
consistent with corrections to the factorization relations arising at $\mathcal{O}(\lqcd/m_c)$.

LHCb measurements of the double differential rate
$\d^2\Gamma(\Lambda_b\to\Lambda_c\ell\bar\nu) / (\d q^2\, \d\cos\theta)$, in
addition to the $q^2$ spectrum, will provide the most differential information
measurable in the massless lepton channels ($\mu$ and $e$), if the details of
the $\Lambda_c$ decay are ignored. Besides the $q^2$ spectrum and the ($q^2$
dependent) forward-backward asymmetry, this double differential distribution
involves a third function of $q^2$, which can help constrain form factors and
test heavy quark symmetry.  If the absolute normalization and the double
differential rate of semileptonic $\Lambda_b \to \Lambda_c$ decays can be
measured, it will provide a fully complementary path to extract $|V_{cb}|$,
explore the $b\to c\tau\nu$ anomalies, and test HQET.  We look forward to these
developments.

\acknowledgments

We thank Marina Artuso, Sheldon Stone, and Mark Wise for helpful conversations.
We thank the Aspen Center of Physics, supported by the NSF grant 
PHY-1607611, where parts of this work were completed. 
FB and WS were supported by the DFG Emmy-Noether Grant No.\ BE~6075/1-1.
ZL was supported in part by the U.S.\ Department of Energy under
contract DE-AC02-05CH11231. 
The work of DR was supported in part by NSF grant PHY-1720252.

\appendix

\section{Form factor definitions, conversions, relations}
\label{sec:FFapp}

The form factors in Eqs.~(\ref{QCDffdef}) and (\ref{HQETffdef}) are related
via~\cite{Falk:1992ws}
\begin{align}
F_1 &= f_1 + (m_{\Lambda_b} + m_{\Lambda_c}) 
  \bigg(\frac{f_2}{2m_{\Lambda_b}} + \frac{f_3}{2m_{\Lambda_c}}\bigg) , \quad
F_2 = -\frac{f_2}{2m_{\Lambda_b}} - \frac{f_3}{2m_{\Lambda_c}}\,,\quad
F_3 = \frac{f_2}{2m_{\Lambda_b}} - \frac{f_3}{2m_{\Lambda_c}}\,, \nn\\*
G_1 &= g_1 - (m_{\Lambda_b} - m_{\Lambda_c}) 
  \bigg(\frac{g_2}{2m_{\Lambda_b}} + \frac{g_3}{2m_{\Lambda_c}}\bigg) ,\quad 
G_2 = -\frac{g_2}{2m_{\Lambda_b}} - \frac{g_3}{2m_{\Lambda_c}} \,,\quad
G_3 = \frac{g_2}{2m_{\Lambda_b}} - \frac{g_3}{2m_{\Lambda_c}} \,,
\end{align}
or in the opposite direction,
\begin{align}\label{qcdtohqet}
f_1 &= F_1 + F_2\, (m_{\Lambda_b} + m_{\Lambda_c})\,,\qquad
  f_2 = (F_3 - F_2)\, m_{\Lambda_b}\,, \qquad
  f_3 = - (F_3 + F_2)\, m_{\Lambda_c}\,, \nn\\*
g_1 &= G_1 - G_2 (m_{\Lambda_b} - m_{\Lambda_c})\,,\qquad
  g_2 = (G_3 - G_2)\, m_{\Lambda_b}\,,\qquad 
  g_3 = -(G_3 + G_2)\, m_{\Lambda_c}\,. 
\end{align}

The form factors used in the lattice QCD calculation~\cite{Detmold:2015aaa} and
in the LHCb analysis~\cite{Aaij:2017svr} follow the definitions in
Ref.~\cite{Feldmann:2011xf},
\begin{align}\label{latticeffdef}
\langle \Lambda_c(p',s') | \bar q\,\gamma_\mu\, b | \Lambda_b(p,s) \rangle
&= \bar u(p',s')\, \bigg[ f_0\, 
  \frac{m_{\Lambda_b}-m_{\Lambda_c}}{q^2}\, q_\mu \nn\\*
&\qquad + f_+\, \frac{m_{\Lambda_b}+m_{\Lambda_c}}{s_+}\, \bigg( p_\mu+p'_\mu
  - \frac{m_{\Lambda_b}^2-m_{\Lambda_c}^2}{q^2}\, q_\mu \bigg) \nn\\*
&\qquad + f_\perp\, \bigg(\gamma_\mu - \frac{2m_{\Lambda_c}}{s_+}\, p_\mu
  - \frac{2m_{\Lambda_b}}{s_+}\, p'_\mu \bigg) \bigg]\, u(p,s)\,, \nn\\
\langle\Lambda_c(p',s') | \bar q\,\gamma_\mu\gamma_5\, b | \Lambda_b(p,s)\rangle
&= -\bar u(p',s')\, \gamma_5\, \bigg[ g_0\,
  \frac{m_{\Lambda_b}+m_{\Lambda_c}}{q^2}\, q_\mu \nn\\*
&\qquad + g_+\, \frac{m_{\Lambda_b}-m_{\Lambda_c}}{s_-}\, \bigg( p_\mu+p'_\mu
  - \frac{m_{\Lambda_b}^2-m_{\Lambda_c}^2}{q^2}\, q_\mu \bigg) \nn\\*
&\qquad + g_\perp\, \bigg(\gamma_\mu + \frac{2m_{\Lambda_c}}{s_-}\, p_\mu 
  - \frac{2 m_{\Lambda_b}}{s_-}\, p'_\mu \bigg) \bigg]\, u(p,s)\,,
\end{align}
where $q=p-p^\prime$, and  $s_\pm = (m_{\Lambda_b} \pm m_{\Lambda_c})^2-q^2 =
2\, m_{\Lambda_b} m_{\Lambda_c} (w \pm 1)$.  These form factors are related to
the HQET form factors defined in Eq.~(\ref{HQETffdef}) via
\begin{align}\label{latticetohqet}
f_1 = f_\perp , \qquad
f_2 &= \frac{f_+ - f_\perp}{w+1} - (f_+ - f_0)\, \frac{1-\rC}\mSqq \,, \nn\\*
f_3 &= \frac{f_+ - f_\perp}{w+1} 
  + (f_+ - f_0)\, \frac{\rC(1-\rC)}\mSqq \,, \nn\\*
g_1 = g_\perp , \qquad
g_2 &= \frac{g_+ - g_\perp}{w-1} + (g_+ - g_0)\, \frac{1+\rC}\mSqq \,, \nn\\*
g_3 &= \frac{g_+ - g_\perp}{w-1} - (g_+ - g_0)\, \frac{\rC(1+\rC)}\mSqq \,.
\end{align}
At $w=1$, corresponding to $q^2_{\rm max}$, the form factors satisfy 
$g_+(q^2_{\rm max}) = g_\perp(q^2_{\rm max})$.

In the heavy quark limit, $f_0 = f_+ = f_\perp = g_0 = g_+ = g_\perp = \zeta +
{\cal O}(\alpha_s,\, \lqcd/m_{c,b})$.
The lattice QCD results in Fig.~12 in Ref.~\cite{Detmold:2015aaa} show that
$f_0\,,\ f_+\,,\ g_0\,,\ g_+\,,\ g_\perp$ differ from one another by less than
${\cal O}(10\%)$, however, $f_\perp$ is substantially enhanced, consistent with
the HQET prediction in Eq.~(\ref{f1g1ratio}).

The form factors in Eq.~\eqref{latticeffdef}, expressed in terms of the
HQET definitions in Eq.~\eqref{HQETffdef}, are
\begin{align}\label{hqettolattice}
f_\perp = f_1\,,\qquad
f_0 &= f_1 + \frac{f_2(1-w\, \rC) + f_3(w -\rC)}{1-\rC}\,, \nn\\*
f_+ &= f_1 + (w+1)\, \frac{f_2\, \rC + f_3}{1+\rC}\,, \nn\\*
g_\perp = g_1\,,\qquad
g_0 &= g_1 - \frac{g_2(1-w\, \rC) + g_3(w-\rC)}{1+\rC}\,, \nn\\*
g_+ &= g_1 - (w-1)\, \frac{g_2\, \rC + g_3}{1-\rC}\,.
\end{align}

\begin{figure}[th]
\includegraphics[width=0.43\textwidth, clip, bb=0 0 420 315]{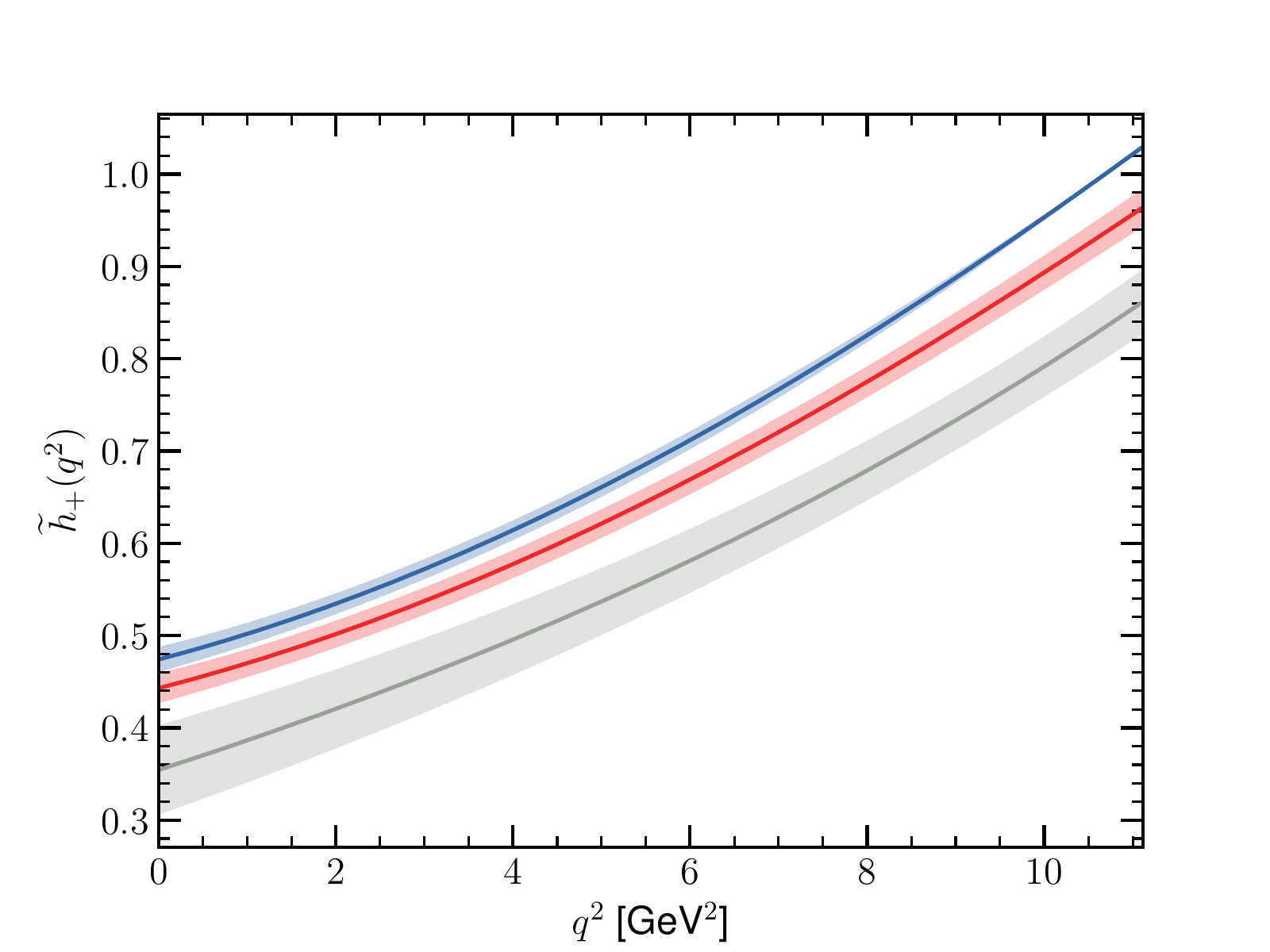}\hfil
\includegraphics[width=0.43\textwidth, clip, bb=0 0 420 315]{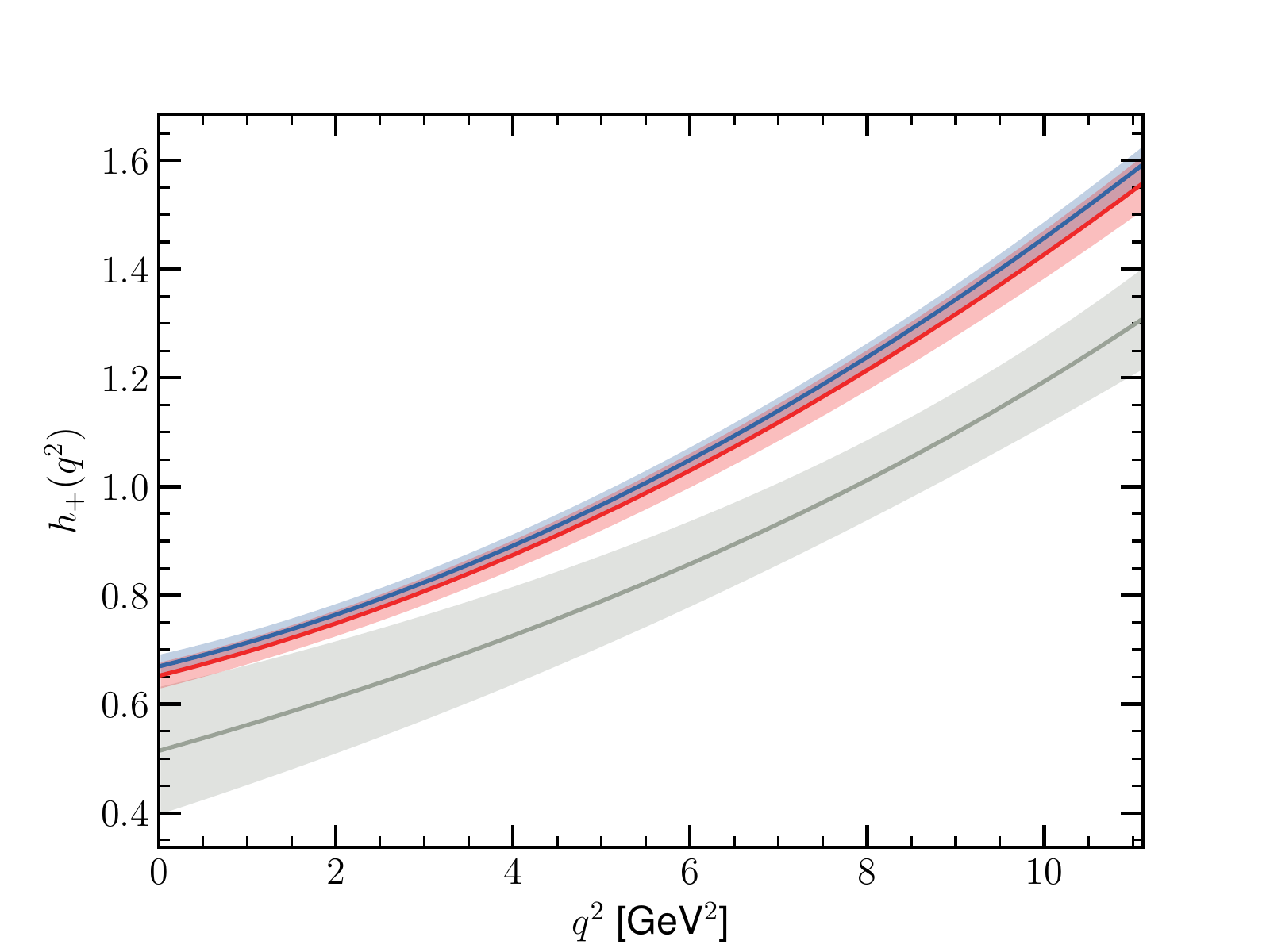}
\\[8pt]
\includegraphics[width=0.43\textwidth, clip, bb=0 0 420 315]{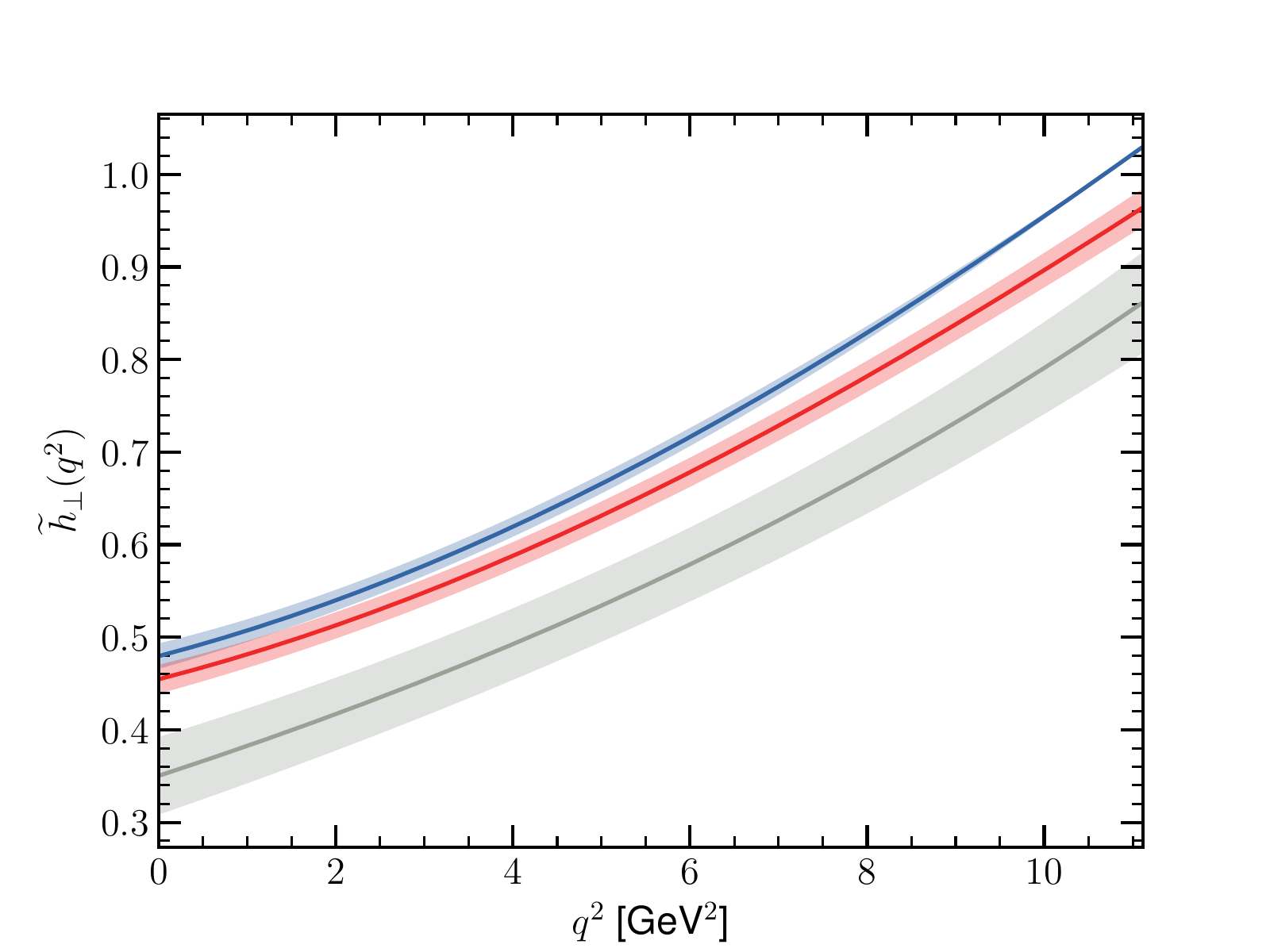}\hfil
\includegraphics[width=0.43\textwidth, clip, bb=0 0 420 315]{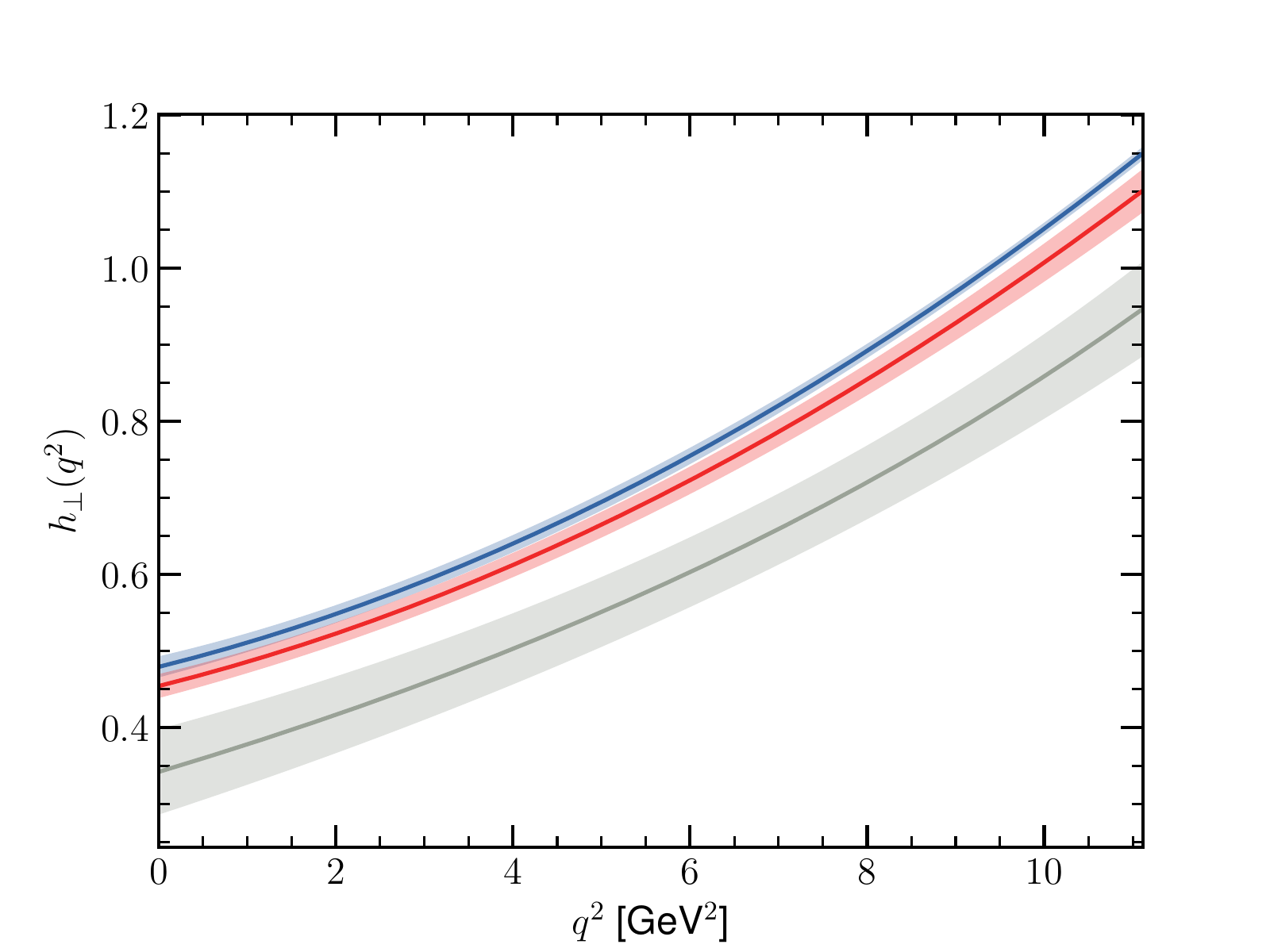}
\caption{Predictions for the tensor form factors in the basis used in the LQCD
calculation~\cite{Datta:2017aue} (scaled to $\mu = \sqrt{m_b m_c}$), compared
with our predictions based on Eq.~(\ref{ffexpbsm}) and the fit to the LHCb data
and the LQCD (axial)vector form factors.  The notation is the same as in
Fig.~\ref{fig:tensorff}.}
\label{fig:tensorffback}
\end{figure}

Finally, the translation between the $h_{1,2,3,4}$ tensor form factors used in
this paper, defined in Eq.~(\ref{HQETffdefnew}), and those defined in Eq.~(2.14)
in Ref.~\cite{Datta:2017aue} are
\begin{align}\label{tensorFFrel}
h_+ &= h_1 - h_2 + h_3 - h_4\, (w+1)\,, \nn\\*
h_\perp &= h_1 - h_2\, \frac{1-w\,\rC}{1+\rC}
  - h_3\, \frac{w-\rC}{1+\rC} \,,\nn\\*
\widetilde h_+ &= h_1 \,, \nn\\*
\widetilde h_\perp &= h_1 - \frac{h_2\, \rC + h_3}{1-\rC}\, (w-1)\,,
\end{align}
and in the opposite direction,
\begin{align}\label{tensorFFrelB}
h_1 &= \widetilde h_+ \,, \nn\\*
h_2 &= \frac{\widetilde h_\perp - \widetilde h_+}{w-1}
  + \big(\widetilde h_\perp - h_\perp\big) \frac{1 + \rC}{\mSqq}\,, \nn\\*
h_3 &= \frac{\widetilde h_+ - \widetilde h_\perp}{w-1}
  + \big(h_\perp - \widetilde h_\perp\big)
  \frac{\rC(1 + \rC)}{\mSqq} \,, \nn\\*
h_4 &= \frac{\widetilde h_+ - \widetilde h_\perp}{w-1}
  + \frac{h_\perp - h_+}{w+1}
  + 2 \big(h_\perp - \widetilde h_\perp\big) \frac{\rC}{\mSqq} \,.
\end{align}
In the heavy quark limit, the tensor form factors calculated in LQCD and shown
in Fig.~2 of Ref.~\cite{Datta:2017aue} satisfy $h_+ = h_\perp = \widetilde h_+ =
\widetilde h_\perp = \zeta + {\cal O}(\alpha_s,\, \lqcd/m_{c,b})$.

\section{Amplitudes}
\label{app:NPampl}

In this appendix we collect explicit expressions for the $\Lambda_b \to
\Lambda_c \ell \nu$ amplitudes,  including mass terms and right-handed sterile neutrino
contributions.  These amplitudes correspond to those used in the \texttt{Hammer}
code~\cite{Duell:2016maj}.

As in Ref.~\cite{Ligeti:2016npd}, we write explicit expressions for the $\bbar \to \cbar$ amplitudes rather than $b \to c$, defining the basis of NP operators to be
\begin{subequations}\label{abdef}
\begin{align}
\text{SM:\,} & \phantom{-}i2\sqrt{2}\, V_{cb}^* G_F\big[\bbar \g^\mu P_L c\big] \big[\bar\nu \g_\mu P_L \ell\big]\,, \\*
\text{Vector:\,} & \phantom{-}i2\sqrt{2}\, V_{cb}^* G_F
  \big[\bbar\big(\alVL \g^\mu P_L + \alVR \g^\mu P_R\big)c\big] 
  \big[\bar\nu\big(\beVL \g_\mu P_L + \beVR \g_\mu P_R\big) \ell\big]\,, \\
\text{Scalar:\,} & -i2\sqrt{2}\, V_{cb}^* G_F
  \big[\bbar\big(\alSL P_L + \alSR P_R\big)c\big]
  \big[\bar\nu\big(\beSL P_R + \beSR P_L\big) \ell\big]\,, \\
\text{Tensor:\,} & -i2\sqrt{2}\, V_{cb}^* G_F
  \big[ \big(\bbar \alTR \sigma^\mn P_R c\big)
  \big(\bar\nu \beTL \sigma_\mn P_R \ell \big)  + \big(\bbar \alTL \sigma^\mn P_L c\big)
  \big(\bar\nu \beTR \sigma_\mn P_L \ell \big) \big]\,.
\end{align}
\end{subequations}
The lower index of $\beta$ denotes the $\nu$ chirality and the lower
index if $\alpha$ is that of the $c$ quark.
Operators for the CP conjugate $b \to c$ processes follow by Hermitian
conjugation. (The correspondence between the $\alpha$, $\beta$ coefficients and the basis typically chosen
for $b\to c$ operators can be found in Ref.~\cite{Bernlochner:2017jxt}.) The $\Lambda_b \to \Lambda_c \ell \nu$ process has four external
spins: $s_b=\pm$, $s_c=1,2$, $s_\ell=1,2$ and $s_\nu=\pm$.  (We label the
$\Lambda_c$ and $\ell$ spin by $1$ and $2$,  to match the conventions of
Ref.~\cite{Ligeti:2016npd} for massive spinors on internal lines.)

Helicity angles and momenta are similarly defined with respect to the $\bbar \to
\cbar$ process. Definitions for the conjugate process follow by replacing all
particles with their antiparticles. The single physical polar helicity angle,
$\thtau$, defines the orientation of the lepton  momenta in their center of mass
reference frame, with respect to $-\bm{p}_{\Lambda_b}$, as shown in Fig.~4 of
Ref.~\cite{Bernlochner:2017jxt}. Note that $\thtau = \pi - \theta$, for $\theta$ defined in Eq.~\eqref{eqn:dGdwdcosth}. 

If subsequent $\Lambda_c \to \Lambda Y$
decays are included coherently, one further defines $\phtau$ and $\phL$ as twist angles of the
$\ell$--$\nu$ and $\Lambda$--$Y$ decay planes, with the combination $\phtau - \phL$
becoming a physical phase. Our phase conventions match the spinor conventions of
Ref.~\cite{Ligeti:2016npd} for not only $\tau$ but also $\Lambda_c$ decay amplitudes.
This amounts to requiring the inclusion in the $\tau$ and/or $\Lambda_c$ decay
amplitudes of an additional spinor phase function, $h_{s_\ell}(s_{\nu})$ and
$h_{s_c}(s_b)$, defined with respect to $s_{\nu}$ and $s_b$, such that
$h_{1}(\dn) = 1 = h_{2}(\up)$, $h_{1}(\up) = e^{i\phtau}$ and $h_{2}(\dn) =
e^{-i\phtau}$. Under these conventions, the $\Lambda_b \to \Lambda_c \ell \nu$ amplitudes themselves are independent of $\phtau - \phL$.

For compact expression of the amplitudes, it is convenient to define 
\begin{equation}
	w_\pm = w \pm \sqrt{w^2-1}\,, \qquad \mSqq = q^2/m_{\Lambda_b}^2 = 1 - 2 \rC w + \rC^2\,, \qquad r_\ell = m_\ell/m_{\Lambda_b}\,,
\end{equation}	
along with
\begin{align}
\WpmP & = \sqrt{w_+} + \sqrt{w_-}\,, & \WpmM & = \sqrt{w_+} - \sqrt{w_-}\,, \nn\\
R_{+\pm} &= (1 + \rC) \pm (1 -\rC)\cos\thtau\,, & R_{-\pm} &= (1 - \rC) \pm (1 + \rC)\cos\thtau\,, \nn\\ 
\Omega_{+} &= r - w + \sqrt{w^2 -1}\, \cos\thtau\,, & \Omega_{\times} &= r w- 1 + r\sqrt{w^2 -1}\, \cos\thtau\,.
\end{align}
The $\Lambda_b \to \Lambda_c \ell \nu$ amplitudes obey the conjugation relation
\begin{equation}
	\label{eqn:parrel}
	\mathcal{A}_{\bar s_b \bar s_c s_\ell s_\nu}\big(w, \sqrt{w^2-1}, \thtau, \phtau\big) = \mathcal{A}_{s_b s_c s_\ell s_\nu}\big(w, -\sqrt{w^2-1}, \pi- \thtau, -\phtau\big)\,,
\end{equation}
in which the exchange $\sqrt{w^2 -1} \to -\sqrt{w^2-1}$ implies also $w_-
\leftrightarrow w_+$. One then need only write the $s_b = -$ amplitudes, with
the $s_b = +$ amplitudes following via Eq.~\eqref{eqn:parrel}. Further writing
$\mathcal{A} = 2\sqrt{2}G_F m_{\Lambda_b}^2\sqrt{\rC(\mSqq -\rl)} \times A$,
the explicit amplitudes are
\begin{subequations}
\begin{align}
A_{\dn 1 1 \dn} & = 
 \bigg\{ -\frac{1}{2} h_S  (\alSL +\alSR ) \beSL  \WpmP+\frac{1}{2} h_P  (\alSL -\alSR ) \beSL  \WpmM \notag \\ 
    & +\frac{f_1  (1+(\alVR+\alVL)\beVL) \rt \big(\sqrt{w_{-}} \RCmp +\sqrt{w_{+}} \RCmm \big)}{2 \mSqq } \notag \\ 
    & -\frac{f_3  (1+(\alVR+\alVL)\beVL) \rt \WpmP \Op }{2 \mSqq }-\frac{f_2  (1+(\alVR+\alVL)\beVL) \rt \WpmP \Ot }{2 \mSqq } \notag \\ 
    & +\frac{g_1  (1+(\alVL-\alVR)\beVL) \rt \big(\sqrt{w_{-}} \RCpp -\sqrt{w_{+}} \RCpm \big)}{2 \mSqq } \notag \\ 
    & -\frac{g_3  (1+(\alVL-\alVR)\beVL) \rt \WpmM \Op }{2 \mSqq }-\frac{g_2  (1+(\alVL-\alVR)\beVL) \rt \WpmM \Ot }{2 \mSqq } \notag \\ 
    & +4 h_1  \alTR  \beTL  \sqrt{w_{+}} \cos\thtau -2 h_2  \alTR  \beTL  \WpmM \cos\thtau  \notag \\ 
    & +2 h_3  \alTR  \beTL  \WpmM \cos\thtau -2 h_4  \alTR  \beTL  (w+1) \WpmM \cos\thtau \bigg\}\\ 
A_{\dn 1 1 \up} & = 
 \sin\thtau \bigg\{ \frac{(1+\rC) f_1  (\alVL +\alVR ) \beVR  \WpmM}{2 \sqrt{\mSqq }} \notag \\ 
    & +\frac{\rC f_2  (\alVL +\alVR ) \beVR  (w+1) \WpmM}{2 \sqrt{\mSqq }}+\frac{f_3  (\alVL +\alVR ) \beVR  (w+1) \WpmM}{2 \sqrt{\mSqq }} \notag \\ 
    & +\frac{(\rC-1) g_1  (\alVL -\alVR ) \beVR  \WpmP}{2 \sqrt{\mSqq }} \notag \\ 
    & +\frac{\rC g_2  (\alVL -\alVR ) \beVR  (w-1) \WpmP}{2 \sqrt{\mSqq }}+\frac{g_3  (\alVL -\alVR ) \beVR  (w-1) \WpmP}{2 \sqrt{\mSqq }} \notag \\ 
    & +4 h_1  \alTL  \beTR  \rt \sqrt{\frac{w_{-}}{\mSqq }}+\frac{2 h_2  \alTL  \beTR  \rt \WpmM}{\sqrt{\mSqq }} \notag \\ 
    & -\frac{2 h_3  \alTL  \beTR  \rt \WpmM}{\sqrt{\mSqq }}+\frac{2 h_4  \alTL  \beTR  \rt (w+1) \WpmM}{\sqrt{\mSqq }}\bigg\}\\ 
A_{\dn 1 2 \dn} & = 
 \sin\thtau \bigg\{ \frac{(1+\rC) f_1  (1+(\alVR+\alVL)\beVL) \WpmM}{2 \sqrt{\mSqq }} \notag \\ 
    & +\frac{\rC f_2  (1+(\alVR+\alVL)\beVL) (w+1) \WpmM}{2 \sqrt{\mSqq }}+\frac{f_3  (1+(\alVR+\alVL)\beVL) (w+1) \WpmM}{2 \sqrt{\mSqq }} \notag \\ 
    & +\frac{(\rC-1) g_1  (1+(\alVL-\alVR)\beVL) \WpmP}{2 \sqrt{\mSqq }} \notag \\ 
    & +\frac{\rC g_2  (1+(\alVL-\alVR)\beVL) (w-1) \WpmP}{2 \sqrt{\mSqq }}+\frac{g_3  (1+(\alVL-\alVR)\beVL) (w-1) \WpmP}{2 \sqrt{\mSqq }} \notag \\ 
    & -4 h_1  \alTR  \beTL  \rt \sqrt{\frac{w_{+}}{\mSqq }}+\frac{2 h_2  \alTR  \beTL  \rt \WpmM}{\sqrt{\mSqq }} \notag \\ 
    & -\frac{2 h_3  \alTR  \beTL  \rt \WpmM}{\sqrt{\mSqq }}+\frac{2 h_4  \alTR  \beTL  \rt (w+1) \WpmM}{\sqrt{\mSqq }}\bigg\}\\ 
A_{\dn 1 2 \up} & = 
 \bigg\{ \frac{1}{2} h_S  (\alSL +\alSR ) \beSR  \WpmP-\frac{1}{2} h_P  (\alSL -\alSR ) \beSR  \WpmM \notag \\ 
    & -\frac{f_1  (\alVL +\alVR ) \beVR  \rt \big(\sqrt{w_{-}} \RCmp +\sqrt{w_{+}} \RCmm \big)}{2 \mSqq } \notag \\ 
    & +\frac{f_3  (\alVL +\alVR ) \beVR  \rt \WpmP \Op }{2 \mSqq }+\frac{f_2  (\alVL +\alVR ) \beVR  \rt \WpmP \Ot }{2 \mSqq } \notag \\ 
    & -\frac{g_1  (\alVL -\alVR ) \beVR  \rt \big(\sqrt{w_{-}} \RCpp -\sqrt{w_{+}} \RCpm \big)}{2 \mSqq } \notag \\ 
    & +\frac{g_3  (\alVL -\alVR ) \beVR  \rt \WpmM \Op }{2 \mSqq }+\frac{g_2  (\alVL -\alVR ) \beVR  \rt \WpmM \Ot }{2 \mSqq } \notag \\ 
    & +4 h_1  \alTL  \beTR  \sqrt{w_{-}} \cos\thtau +2 h_2  \alTL  \beTR  \WpmM \cos\thtau  \notag \\ 
    & -2 h_3  \alTL  \beTR  \WpmM \cos\thtau +2 h_4  \alTL  \beTR  (w+1) \WpmM \cos\thtau \bigg\}\\ 
A_{\dn 2 1 \dn} & = 
 \sin\thtau \bigg\{ \frac{f_1  (1+(\alVR+\alVL)\beVL) \rt \WpmM}{2 \sqrt{\mSqq }} \notag \\ 
    & +\frac{g_1  (1+(\alVL-\alVR)\beVL) \rt \WpmP}{2 \sqrt{\mSqq }} \notag \\ 
    & +\frac{4 h_1  \alTR  \beTL  (w_{-} - \rC)}{\sqrt{\mSqq  w_{-}}}-\frac{2 h_2  \alTR  \beTL  (\rC w_{+} - 1) \WpmM}{\sqrt{\mSqq }} \notag \\ 
    & -\frac{2 h_3  \alTR  \beTL  (\rC-w_{-}) \WpmM}{\sqrt{\mSqq }}\bigg\}\\ 
A_{\dn 2 1 \up} & = 
 \sin^2\frac{\thtau}{2}\bigg\{ -f_1  (\alVL +\alVR ) \beVR  \WpmM \notag \\ 
    & +g_1  (-\alVL +\alVR ) \beVR  \WpmP \notag \\ 
    & -\frac{8 h_1  \alTL  \beTR  \rt \sqrt{w_{+}} (\rC w_{-} - 1)}{\mSqq }+\frac{4 h_2  \alTL  \beTR  \rt (\rC w_{-} - 1) \WpmM}{\mSqq } \notag \\ 
    & -\frac{4 h_3  \alTL  \beTR  \rt (w_{+} - \rC) \WpmM}{\mSqq }\bigg\}\\ 
A_{\dn 2 2 \dn} & = 
 \cos^2\frac{\thtau}{2}\bigg\{ f_1  (1+(\alVR+\alVL)\beVL) \WpmM \notag \\ 
    & +g_1  (1+(\alVL-\alVR)\beVL) \WpmP \notag \\ 
    & -\frac{8 h_1  \alTR  \beTL  \rt (\rC w_{+} - 1) \sqrt{w_{-}}}{\mSqq }-\frac{4 h_2  \alTR  \beTL  \rt (\rC w_{+} - 1) \WpmM}{\mSqq } \notag \\ 
    & -\frac{4 h_3  \alTR  \beTL  \rt (\rC-w_{-}) \WpmM}{\mSqq }\bigg\}\\ 
A_{\dn 2 2 \up} & = 
 \sin\thtau \bigg\{ -\frac{f_1  (\alVL +\alVR ) \beVR  \rt \WpmM}{2 \sqrt{\mSqq }} \notag \\ 
    & +\frac{g_1  (-\alVL +\alVR ) \beVR  \rt \WpmP}{2 \sqrt{\mSqq }} \notag \\ 
    & +\frac{4 h_1  \alTL  \beTR  (w_{+} - \rC)}{\sqrt{\mSqq  w_{+}}}+\frac{2 h_2  \alTL  \beTR  (\rC w_{-} - 1) \WpmM}{\sqrt{\mSqq }} \notag \\ 
    & -\frac{2 h_3  \alTL  \beTR  (w_{+} - \rC) \WpmM}{\sqrt{\mSqq }}\bigg\}\,.
\end{align}
\end{subequations}
The total differential rate for $\Lambda_b \to \Lambda_c \ell \nu$ is obtained from these expressions via 
\begin{equation}
	d\Gamma = \frac{G_F^2 m_{\Lambda_b}^5 \rC^3 |V_{cb}|^2}{32\pi^3} \sqrt{w^2-1} \frac{(\mSqq - \rl)^2}{\mSqq} \sum_{s_b, s_c, s_\ell, s_\nu} |A_{s_b, s_c, s_\ell, s_\nu} |^2 dw \,\sin\thtau d\thtau\,.
\end{equation}

\end{document}